\documentclass[12pt,journal,compsoc]{IEEEtran}
\usepackage{multirow}
\usepackage{booktabs}
\usepackage{graphicx}
\usepackage{pbox}
\usepackage{array}
\usepackage[english]{babel}
\usepackage{float}

\usepackage{hyperref} 
\begin{document}
\title{A simple agent-based spatial model of the economy:  tools for policy}
\author{Bernardo~Alves~Furtado,~{DISET/IPEA,~CNPq,}
        and~Isaque~Daniel~Rocha~Eberhardt,~{DISET/IPEA,~PhD~candidate~UnB,}%
\IEEEcompsocitemizethanks{\IEEEcompsocthanksitem B. A. Furtado is with the Department of Innovation and Production Studies at the Institute for Applied Economic Research and with the National Council of Research (Brazil).\protect\\
E-mail: bernardo.furtado@ipea.gov.br
\IEEEcompsocthanksitem I. D. R. Eberhardt is with the Institute for Applied Economic Research and the Department of Transport at University of Bras\'ilia}%
\thanks{Manuscript reviewed October 6, 2016.}}

\IEEEtitleabstractindextext{
\begin{abstract}
This study simulates the evolution of artificial economies in order to understand the tax relevance of administrative boundaries in the quality of life of its citizens. The modeling involves the construction of a computational algorithm, which includes citizens, bounded into families; firms and governments; all of them interacting in markets for goods, labor and real estate. The real estate market allows families to move to dwellings with higher quality or lower price when the families capitalize property values. The goods market allows consumers to search on a flexible number of firms choosing by price and proximity. The labor market entails a matching process between firms (given its location) and candidates, according to their qualification. The government may be configured into one, four or seven distinct sub-national governments, which are all economically conurbated. The role of government is to collect taxes on the value added of firms in its territory and invest the taxes into higher levels of quality of life for residents. The model does not have a credit market, given the emphasis of the research question on the relevance of municipal administrative boundaries. The analysis of the markets indicate development paths and data-generating mechanisms for each territorial approach used. The results suggest that the configuration of administrative boundaries is relevant to the levels of quality of life arising from the reversal of taxes. The model with seven regions is more dynamic, with higher GDP values, but more unequal and heterogeneous across regions. The simulation with only one region is more homogeneously poor. The study seeks to contribute to a theoretical and methodological framework as well as to describe, operationalize and test computer models of public finance analysis, with explicitly spatial and dynamic emphasis. Several alternatives of expansion of the model for future research are described. Moreover, this study adds to the existing literature in the realm of simple microeconomic computational models, specifying structural relationships between local governments and firms, consumers and dwellings mediated by distance.
\end{abstract}

}

\maketitle
\IEEEdisplaynontitleabstractindextext
\IEEEpeerreviewmaketitle

\section{Introduction and Literature}

\IEEEPARstart{T}{he} Brazilian tax system is paradoxical, with high taxes, dual tax systems (taxes and contributions) and fierce fiscal war between federated members \cite{afonso_public_2013}. The complexity of the tax system becomes more obvious and striking when considering the subnational entities. The post-1988 constitutional decentralization imposes the same competences to very heterogeneous municipalities \cite{rezende_federalismo_2010}. Municipalities that have different administrative, technical, and political capacities; besides their inherently differentiated borrowing leverage \cite{canuto_until_2013}. This heterogeneity among municipalities occurs not only in relation to budgetary magnitude, but also with regard to the disparity between central and peripheral municipalities in metropolitan and regional context \cite{antinarelli_federalismo_2012,rezende_financing_2006}. Indeed, Furtado et al. \cite{furtado_fatos_2013} identified that there are significantly fewer resources to metropolitan peripheral municipalities vis-\`a-vis the central city and non-metropolitan municipalities. The authors also suggest that such municipalities are administratively inefficient, with higher expenditures and poorer results. In addition to this reduced administrative and financial capacity, peripheral municipalities still have worse quality of life and higher levels of violence \cite{andrade_vulnerabilidade_2005,waiselfisz_mapa_2012}. There is a huge amount of literature on public spending efficiency \cite{afonso_public_2013,gasparini_transferencias_2011,orair_uma_2011}, which contains actual policy propositions \cite{afonso_imposto_2014,gobetti_ajuste_2015}, is descriptive \cite{santos_financas_2014}, and of a high-quality level. However, few exercises emphasize the prospective analysis that simulates future effects of present public policy change \cite{brandalise_simulacao_2012,carvalho_cenarios_2015}, especially for the case of Brazil and its subnational entities.

Computer simulation models for macroeconomic analysis and taxes \cite{dawid_labor_2012,dosi_income_2012,dosi_microfoundations_2009}, banking and finance \cite{cajueiro_possible_2005,cajueiro_role_2008,tabak_topological_2009}, stock exchange \cite{lebaron_agent-based_2006,palmer_artificial_1994} and energy market \cite{lebaron_modeling_2008}, to name a few applications, abound. These studies were developed from the seminal works of Anderson, Pines and Arrow \cite{anderson_economy_1988} and Arthur \cite{arthur_inductive_1994}. Recently, advances in this literature includes models that discusses bank interconnections by means of network analysis and systemic fault possibilities \cite{bargigli_interaction_2014,grilli_markets_2014,ya-qi_rumor_2013}

This abundant literature, however, looks at specific markets (banking, energy or exchange markets) or seek to represent markets and its agents and processes in detail, so that they quickly become complex and demanding high computing power \cite{van_der_hoog_production_2008,guocheng_application_2015}.

Simple models that intend to model the interaction among actors in short-term spatial scales are rare. Tesfatsion \cite{tesfatsion_agent-based_2006} makes an initial proposal of a model with two products (hash and beans) whereas Straatman et al. \cite{straatman_generic_2013} proposed a framework that simulates a market auctions linked to a production model that together result in a simple model, but complete and micro founded.

Lengnick \cite{lengnick_agent-based_2013} expands the work of Gaffeo et al. \cite{gaffeo_adaptive_2008} and proposes a model that simulates macroeconomic variables, contains elements of real estate and goods and labor markets. As detailed below, our proposal is based on Lengnick's model, but makes several changes, including explicit space in the housing market, and subnational administrative regions.

Given this framework, this paper proposes an agent-based model that is able to replicate basic elements of an economy, its markets, its players and its processes as simple as possible, enabling spatial and dynamic analysis of the central economic mechanisms. The intention is to understand the mechanisms that generates the observed data, so that prospective analysis can be made. Specifically, the research question is to identify whether the change of administrative boundaries and the consequent change of local tax revenue dynamics, in principle, alters the quality of life of the citizens.

In addition to answering the research question, the contribution of this study is the explicit construction of a computational algorithm that can be configured as a 'simulation engine of the economy'. The paper can be said to be a modular laboratory on which small changes and additions can be applied in order to amplify research possibilities. Thus, the fourth section includes specific examples of future applications of the model in addition to the exercise done in this text.

The model adds to the literature as an adaptation and advancement of the approach proposed by Lengnick \cite{lengnick_agent-based_2013}. The main contribution is the inclusion of local governments to collect taxes and provide public services. However, the proposed model has a different objective from the original. Whereas Lengnick seeks to study effects on macroeconomic variables of small shocks of monetary policy, this model emphasizes the spatial differences among different administrative regions that collect taxes and invest in their own regions through public service provision, hence promoting the improvement of quality of life of local people. Moreover, the design of the model is also innovative, changing a fixed dwelling structure into one in which families move in search of homes and regions either with better quality of life – \`a la Tiebout \cite{tiebout_pure_1956} – or that best suits their current income status. \footnote{See Pinto's \cite[p. 75]{pinto_direito_2014} discussion: "The decentralization and fragmentation of the territory poses alternatives for consumers of collective services who cannot buy services individually, but may buy a package of services and goods that are more preferable".}

Another important distinction of our model is the absence of a network-like structure that establishes the interactions of the labor market and the goods market. In our model the interactions in these markets take place through prices and the distance between the dwelling and the firm. Finally, the entry into the labor market is restricted only by age and open to all members of the family, whereas in Lengnick \cite{lengnick_agent-based_2013} it is exclusive of the head of the family.

Hence, this paper proposes a simulation model of the economy which is based on previous literature, but advances in the specific area of simple microeconomic models, introducing local governments and explicit spatiality of markets.

Besides this introduction, the text includes the presentation of the model (\autoref{sec2}), followed by the presentation and discussion of the results (\autoref{sec3}) and the proper sensitivity analysis. \autoref{sec4} specifies possibilities of further development of the model that can be easily applied, justifying the work as a theoretical and methodological 'proposal'. \autoref{sec5} presents the final considerations.

\section{The proposed model: methodology, features and processes}\label{sec2}

In order to model the collection of local taxes and the provision of public goods to evaluate and compare policy options an agent-based model of a simple economy is presented. We propose a model with heterogeneous agents, dwellings, firms and governments, each with attributes, location and specific processes attached. After the description of the theoretical model, a numerical simulation is applied to the set of parameters, its robustness is verified by a sensitivity analysis and the results for specific periods of time are computed.

\subsection{Agent-based modeling}

\subsubsection{Literature}
The economic analysis based on agent-based models has its methodological groundwork laid by the 'Sugarscape' model, developed by Epstein and Axtell \cite{epstein_growing_1996}. Before that, agent-based models were discussed in the context of social segregation in the classical work of Nobel author Thomas Schelling \cite{schelling_models_1969}; on the seminal framework of game theory and cooperation strategies \cite{axelrod_evolution_1981} and social \cite{holland_complex_1992} and economic sciences \cite{ciarli_structural_2012, holland_artificial_1991}. Furthermore, complete microsimulations models of the labor market were reported much earlier by Bergmann \cite{bergmann_microsimulation_1974} and Eliasson \protect\cite{eliasson_micro-macro_1976}.

More recently, agent-based models have been applied to learning and behavior studies, coalition and cooperation \cite{nardin_trust-based_2012}; artificial intentionality \cite{adamatti_analysis_2009} and education and cognition \cite{maroulis_agent-based_2010,maroulis_modeling_2014}\footnote{For a more detailed review, see Winikoff et al. \cite{winikoff_principles_2012}}. A recent milestone in economics is the text of Boero et al. \cite{boero_agent-based_2015} which offers a conceptual and methodological description, along with applications for human capital development, network analysis, the interbank payment systems, consulting firms, insurance systems in health, ex-ante evaluation of public policies, governance, tax, and cooperation.

The methodological steps of ABMs with an emphasis on interpretation of empirical data  are described by Hassan et al. \cite{hassan_injecting_2010}. Two central aspects of the methodology are verification and validation \cite{carley_validating_1996,midgley_building_2007}. The verification step assesses whether the adopted algorithm effectively does what the modeler and the developer planned. That is, it checks the adequacy of the intention of the algorithm against its factual implementation \cite{david_logic_2005}.

The validation process refers to the use of historical data to assess whether the model can minimally replicate known trajectories. It verifies that the model contains the essence of the phenomenon. Once validated, the model can be used to indicate future trajectories. Zhang et al. \cite{zhang_study_2011} illustrate this process for the adoption of alternative car fuels.

One methodological principle of this modeling process is that the decisions made and the steps of the model are known, understood and comparable. The scientific community suggests two procedures (a) the adoption of protocols, such as the Overview, Design concepts, and Details protocol (ODD), described by Grimm et al. \cite{grimm_standard_2006,grimm_odd_2010}; and (b) the availability of the source code. The code used in this study is available and can be requested to the authors.\footnote{Upon publication, the code  will be made available in GitHub and OpenABM.} The PseudoCodes are available as appendices. The ODD protocol is also available as Appendix \autoref{odd} of this work. 

\subsection{Attributes of the model: processes and rules}

This section describes the model, its characteristics, assumptions, processes, steps, intentions and limitations. Intuitively, we describe the decision-making processes that govern the dynamics of the model. The literature that underlies the choices are listed in the processes. 

\subsubsection{Classes}
The model was developed using the concept of object-oriented programming (OOP) in Python, version 3.4.4. \footnote{For an introduction in Python, see Downey \cite{downey_think_2012}} 

The following section describes the initial values and allocation processes; the breakdown of markets, the government, the spatial and temporal sequencing of the model. Then, we present the implementation, parameters and limitations of the model.

\textit{Classes - initial values} 

The model contains five main classes: agents: citizens, bounded into families; dwellings, firms, and governments. The agents' features are drawn from a uniform distribution and includes age, years of schooling (qualification) and an initial monetary amount.

Dwellings have different sizes. Their prices are a function of its size and the value of the square meter given by its location. Firms are also located randomly in space and start the simulation with some capital. 

\textit{Allocation of agents into families}

The modeler determines the number of agents and the number of families of the model exogenously. The allocation process is random. An agent who has not been allocated is chosen along with a family and the link is made. Thus, the proportion of agents per family is endogenous and variable. Agents maintain the same age throughout the simulation. 

\textit{Initial allocation of families into dwellings}

Before simulation begins, families are randomly allocated to dwellings that are vacant. 

\subsubsection{Real estate market}

When the simulation is already underway, the process of modeling the real estate market is as follows:
Given a parameter chosen by the modeler, say 0.07, that portion of the set of families monthly enters into a randomly composed list of  'families on the market' in pursuit of new residence \cite{arnott_economic_1987}. At the same time, vacant dwellings are selected.\footnote{The number of dwellings should always be larger than the number of families, for this version of the model.} Residential prices $P_{t,i}$ are monthly updated, given the price in the previous month $P_{t-1,i}$ and the percentage of change in the Quality of Life Index $\Delta IQV_r$ of the region where the residence is located.
\begin{equation}
P_{i,t}=P_{i,t-1}*(1+\frac{\Delta IQV_{r}}{IQV_{r,t-1}})
\end{equation}			
The quality of the dwelling $Q_i$ is based on the size of the residence $s_i$ – which is a fixed value – and on the current Quality of Life Index \cite{dipasquale_urban_1996,nadalin_tres_2010} and it serve solely as a choice criteria for the new residence.\footnote{That is, properties' prices are defined by their own features, plus local attributes.} 
\begin{equation}
Q_i=s_i*IQV_{r,t}
\end{equation} 
Two alternatives are available for families who are on the market. Families whose total financial resources is higher than the median of all families will look for houses with higher quality and will conclude the purchase if the value of the current family home $P_{i(s,r)}$ added to the cash available $Y$ is higher than the value of the better quality house intended $P_{i(s,r)}+Y>P_{j(s,r)}$. On the other hand, families whose available resources are less than the median of families' wealth will look for cheaper homes $P_{i(s,r)}>P_{j(s,r)}$, so that they acquire new cash  \cite{brueckner_structure_1987,dipasquale_urban_1996}. 

When observed the conditions, the change of address is made and the difference, if moving into more expensive homes $P_{i(s,r)}-P_{j(s,r)}$, or payback $P_{j(s,r)}-P_{i(s,r)}$ is computed on the family budget. The houses whose families moved become vacant. 

Thus, a portion of the families are always looking for larger or better quality homes, located in the best areas, when they have the financial resources and the other portion of families are in search of cheaper homes from which they can capitalize.

\subsubsection{Firms: production function and prices}

The firm's production technology is fixed and the production function depends on the number of workers $l_f$, their qualification $E_k$ and an exogenous parameter $\alpha $ that determines productivity.\footnote{Adapted from \cite{lengnick_agent-based_2013, dosi_microfoundations_2009, gaffeo_adaptive_2008}.} 
\begin{equation}
f(l_f,E_k,\alpha)=\Sigma l_f*E_k^\alpha
\end{equation}
The production is updated daily according to the above equation. In this model there is only one product per firm. 

\textit{Firms: decision-making about price adjustment}

The literature confirms the rigidity of prices and the difficulty of managerial decision-making about the process of changing prices \cite{blinder_inventories_1982,blinder_sticky_1994}. In the proposed model, the initial price is set as the cost price. Firms change their prices according to inventory levels \cite{bergmann_microsimulation_1974}.\footnote{Bergmann \cite{bergmann_microsimulation_1974} uses cost and profit information in addition to stock levels to determine price changes. Dawid et al. \cite{dawid_labor_2012} use stock levels to define production quantity.}  When the level in stock $q$ is below the level given by the exogenous parameter $\delta$, prices are adjusted upwards, in the amount stipulated by another chosen parameter $\phi$. This parameter is exactly the mark-up chosen by the firm. When the amount is above the chosen level, prices go back to cost price. That is, when demand is low the mark-up is zero. This proposal follows the survey results, conducted by Blinder \cite{blinder_sticky_1994}, which indicates that only a small portion of firms readjusts prices downwards.
\begin{equation}
\left\{\begin{array}{ll}
\mbox{ se } q<\delta , p_t = p_{t-1}*(1+\phi) \\
\mbox{ se } q>\delta , p_t = 1 
\end{array}\right.
\end{equation}
\subsubsection{Goods market}

Given that not all family agents are part of the working population, the family's total resources are equally divided among family members before the decision to consume. Each customer then chooses a value for consumption ranging between 0 and their total wealth $w_i$, discounted by an exogenous factor of propensity to consume $\beta $ \cite{schettini_alises_2012}.
\begin{equation}
C_i=(0,w_i^\beta)
\end{equation}
The family then carries out two calculations. Given the market size parameter $\Gamma$, for example, of five firms, each agent searches among these firms, the one with lowest price \cite{mankiw_principles_2011}, and the one with the shortest distance of the agent's residence  \cite{fujita_spatial_1999,losch_economics_1954}. Randomly, the agent chooses between lower price and shorter distance. Intuitively, sometimes it is worth the effort to go further to find the lowest price, sometimes one chooses closer, though not necessarily cheaper.
\subsubsection{Labor market}

\textit{Wages}

Wages are defined as a fixed portion $k$ \footnote{For this specification of the model, $k$ was set to 0.65} multiplied by the employee qualification $E_i$ elevated to a parameter of productivity $\alpha$. The parameter $\alpha$ is the same parameter of the company production function. This decision is harmonious to the fact that more skilled workers also produce more (in the proposed model). 
\begin{equation}
w_i=k*E_i^\alpha
\end{equation}
Thus, better-qualified (and more productive) employees have better pay.

\textit{Labor market}

The firm makes decisions regarding hiring and firing randomly, on average, once every four months, according to an exogenously set parameter $γ$.
 
The selection is made through a public advertising system. Interested companies become part of a list. Employees between 17 and 70, who are currently not employed, repeatedly, apply themselves to the labor supply list.

Then, there is a matching system between company and employee, so that the randomly selected firm chooses the most qualified employee or the one who lives the closest \cite{boudreau_stratification_2010}. Once the matching has been made, the firm and the hired employee are removed from the list of public announcements and a new round of wage, distances and qualifications ranking is made. And so on, until there is no more interested firms or available employees.\footnote{Neugart et al. \cite{neugart_agent-based_2012} have discussed matching mechanisms for the labor market bu ponder that there is insofar no concensus about the best procedure.} 
When making firing decisions, the firm just randomly chooses an employee and let him or her go.

\subsubsection{Government}

Local governments in each region collect a tax on consumption $\tau$, at the time of purchase in accordance with the location of the firm conducting the sale. The rate is determined by an exogenous parameter.
Every month, the governments of each region $r$ completely transform the resources per capita  collected in linear increases in the Quality of Life Index \cite{schettini_alises_2012}. That is, the QLI is a linear result of the summed sales of firms in a given region, weighted by (ever-changing) population dynamics ($N_r$). 
\begin{equation}
IQV_{r,t}=IQV_{r,t-1}+\Sigma \frac{\tau}{N_r}
\end{equation}
In the model proposed in this paper, three alternatives of government administrative designs are proposed. They are detailed in item 2.3.
\subsubsection{Model sequence}

The model follows the temporal distribution proposed by Lengnick \cite{lengnick_agent-based_2013}, which consists of 21 days to make a month, months are added to quarters and then to years. The sequence of actions occurs with the simultaneous interaction of various classes (see \autoref{tab1}). 

The model sequence can be described as follows:
\begin{enumerate}
\item The modeler defines whether the system should be configured with one, 4 or 7 regions. The simulation parameters (see section 2.4) and the run parameters can be changed.
\item Regions, agents, families, households and firms are created, given the parameters provided.
\item Agents are allocated to families and families are allocated to dwellings. 
\item Before the actual start of the simulation time, the initial framework includes the creation of one product by firm and an initial round of hiring. 
\item When the simulation begins, the production function is applied everyday for all firms.
\item At the end of each month:
\begin{enumerate}
\item Firms pay wages;
\item Households consume and (in the same transaction) governments collect taxes; 
\item Governments apply their available resources into the update of QLI; 
\item Firms update their profits, given their last quarter capital;
\item Firms update product prices; 
\item If necessary, firms post job offers or fire employees; 
\item Unemployed workers offer themselves for the vacancies and the matching process is carried out; 
\item A share of the families enter the housing market and perform transactions. 
\end{enumerate}
\item Every quarter, companies report profits for the period
\end{enumerate}

\begin{table*}[htbp]
  \centering
  \caption{Sequencing and interaction between classes and temporal dynamics of the algorithm. Start processes at 'setup' and 'day 0', followed by days that add up into months and months into quarters, successively, until the period determined by the modeler. Items with an asterisk indicate the need for exogenous parameters}
    \begin{tabular}{r|p{3cm}p{3cm}p{3cm}p{3cm}cc}
    \toprule
    \textbf{} & \textbf{Agents} & \textbf{Families} & \textbf{Firms} & \textbf{Dwellings} & \textbf{Government} \\
    \midrule
    \textbf{Setup} & Creation * &   &   &   &  \\
    \textbf{} &   &   &   &   & Creation * \\
    \textbf{} &   & Creation * &   &   &  \\
    \textbf{} &   &   &   & Creation * &  \\
    \textbf{} &   &   & Creation * &   &  \\
    \textbf{} & Allocation of agents into families & Allocation of agents into families &   &   &  \\
    \textbf{} &   & Allocation families in dwellings. &   & Allocation families in dwellings. &  \\
    \textbf{} &   &   &   &   & Initialization \\
    \hline
    \textbf{Day 0} &   &   & Creation of product &   &  \\
    \textbf{} &   &   & Offer position &   &  \\
    \textbf{} & Apply for position &   &   &   &  \\
    \textbf{} &   &   & Hire * &   &  \\
    \textbf{} &   &   & Address register &   & Register \\
    \textbf{} & Production &   & Production &   &  \\
    \textbf{} &   &   &   & Address register & Register \\
    \textbf{} &   &   &   &   &  \\
    \hline
    \textbf{Days $\Downarrow$} & Production &   & Production &   &  \\
    \hline
    \textbf{Months $\Downarrow$} & Wages &   & Wages &   &  \\
    \textbf{} &   & Family per capita distribution &   &   &  \\
    \textbf{} & Purchase ** &   & Purchase  &   & Purchase (Tax) * \\
    \textbf{} &   &   &   &   & Update QLI \\
    \textbf{} &   &   & Update profits &   &  \\
    \textbf{} &   &   & Decide on prices &   &  \\
    \textbf{} & Fire &   & Offer position/Fire &   &  \\
    \textbf{} & Apply for position &   &   &   &  \\
    \textbf{} &   &   & Hire * &   &  \\
    \textbf{} &   & Enters Real estate market &   & Update prices & Inform QLI \\
    \textbf{} &   &   &   &   &  \\
    \hline
    \textbf{Quarters $\Downarrow$} &   &   & Update profits &   &  \\
    \hline
    \textbf{Years $\Downarrow$} &   &   &   &   &  \\
    \bottomrule
    \end{tabular}%
  \label{tab1}%
\end{table*}%

\subsubsection{Indicators and iterations}

For this paper, the results were obtained with 1,000 iterations for each spatial divisions (one, 4 and 7 regions). 

\subsection{Spatial emphasis of the model}

The model has a clear emphasis on its spatial aspects, as space is central to answer the research question. Calculation of the distance (and accessibility) is present in two moments: (a) at the choice of the employee by the firm and (b) when the consumer chooses between price and (easy) location. As a share of families is relocating, these distance calculations are dynamic and change the relations among firms and consumers and firms and workers every month. That is, firms and dwellings are fixed, but families move constantly, ensuring the spatial dynamics of the model.

Furthermore, the QLI is a linear and spatially compartmentalized reflection of firms’ sales in each region. This same QLI, in turn, affects the prices of dwellings. Thus, the housing market and the goods and labor markets are all spatially linked.

\begin{figure}[!t]
\centering
\includegraphics[width=9cm]{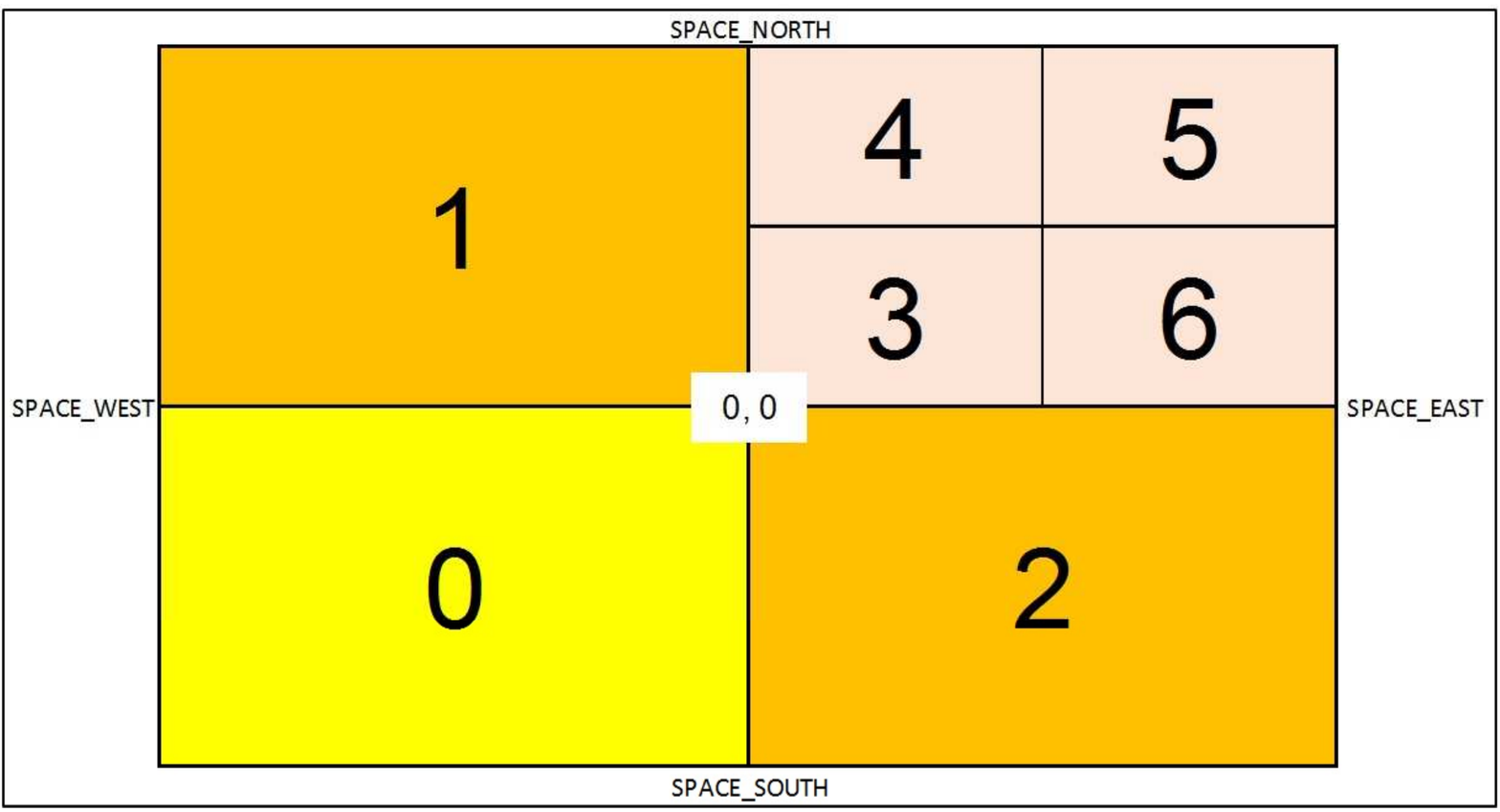}

\caption{Division of space into regions, with coordinates and codes. When running with one region, the simulation space is complete, from north to south, from east to west, including all colors. When running with four regions, the regions are symmetrical, including the codes 0, 1, 2 and the entire region that consists of subregions 3, 4, 5, and 6. With seven, there are larger (0, 1 and 2) and smaller (3, 4, 5 and 6) regions.}
\label{fig1}
\end{figure}

Besides the presence of spatial interaction in the processes themselves, the model also differentiates the applied regions’ design, according to the scheme of \autoref{fig1}. The figure shows the coordinates of the central point (0, 0). Along the four directions, the boundaries can be established by parameters. This study uses the parameters 10, -10, 10, -10, for the North, South, East and West directions, respectively.

Three different design were used and are applied by changing the number of regions $\eta$. If $\eta$ is equal to 1, the model runs with only one region, with code 0, which encompasses the entire space. With $\eta$ equal to 4, the model runs with four regions, with codes 0, 1, 2 and 3, and region 3 covers the entire area of  subregions 3, 4, 5 and 6 in \autoref{fig1}. And finally, with seven regions, the model follows the configuration of the codes of \autoref{fig1}, with four smaller regions and three larger ones.

\subsection{Model implementation and parameters}

Running the model is simple and done with just one command. Optionally, the modeler can set the parameters for each simulation and for the run itself, as described below. A systematic analysis of the parameters that assesses whether small changes significantly affect the results and seek to confirm the robustness of the model (sensitivity analysis) is made after the presentation of results.

For each simulation run you can choose the number of agents, families, households, the time duration in days, the number of local governments in which space is divided (namely, municipalities, with competence over their territory) and the path file to save the results (\autoref{tab2}).

\begin{table*}[htbp]
  \centering
  \caption{Parameters of the simulation (that define each run) and exogenous parameters of the model.}
    \begin{tabular}{p{4cm}p{1cm}p{1cm}p{8.5cm}}
    \toprule
    \multicolumn{1}{l}{\textbf{Simulation parameters}} & \multicolumn{1}{l}{\textbf{Values}} & \multicolumn{1}{l}{\textbf{Possibilities' intervals }} & \multicolumn{1}{l}{\textbf{Observations}} \\
    \midrule
    Number of days & \multicolumn{1}{c}{5,040} & \multicolumn{1}{c}{(63, 12,800)} & The model was developed to run up to 50 years, however with loss of explantory power. We ran the model for 20 years. \\
    Number of agents & \multicolumn{1}{c}{1,000} & \multicolumn{1}{c}{(10, 10,000)} & The growth of the number of agents makes the simulation slower. \\
    Number of families & \multicolumn{1}{c}{400} & \multicolumn{1}{c}{(4, 2,000)} & Endogenously, it's used to define the average number of agents per families. The suggestion is to have 2.5 agents per family, on average \\
    Number of dwellings & \multicolumn{1}{c}{440} & \multicolumn{1}{c}{(5, 2,200)} & Necessarily higher than the number of families number. Vacancy in Brazil is around 11\%. \\
    Number of firms & \multicolumn{1}{c}{110} & \multicolumn{1}{c}{(2, 1,000)} & Approximately 10\% of the number of agents. \\
    Number of regions & \multicolumn{1}{c}{1; 4; 7} & \multicolumn{1}{c}{(1; 4; 7)} & Alternative number of regions to run the model. \\
    \hline
    \textbf{Model parameters} & \multicolumn{1}{c}{} & \multicolumn{1}{c}{} &  \\
    \hline
    \textbf{Firms} & \multicolumn{1}{c}{} & \multicolumn{1}{c}{} &  \\
    Alpha & \multicolumn{1}{c}{0.25} & \multicolumn{1}{c}{(0, 1)} & Production function expoent. When set to "1", it does not change the model, when set to "0", the production of the firm is one unit. \\
    Beta & \multicolumn{1}{c}{0.87} & \multicolumn{1}{c}{(0, 1)} & Consumption function exponent. When set to "1" consumption vary from zero to the total of available money. \\
    Quantity to change prices $\delta$ & \multicolumn{1}{c}{10} & \multicolumn{1}{c}{(100, 2,000)} & Threshold to change prices. \\
    Frequency of entry in the labor market & \multicolumn{1}{c}{0.28} & \multicolumn{1}{c}{(0, 1)} & Time frequency of decision-making on labor market. When set to "0", the evaluation is made every month. When set to "0.25", the firm enters the market three times every four months, on average.\\
    Mark-up & \multicolumn{1}{c}{0.03} & \multicolumn{1}{c}{(0, 1)} & Percentage added to prices when demand  is high (product level on inventory is below "Quantity to change prices").\\
    \textbf{Agents} & \multicolumn{1}{c}{} & \multicolumn{1}{c}{} &  \\
    Market size $\Gamma$ & \multicolumn{1}{c}{100} & \multicolumn{1}{c}{(1, 1,000)} & Number of firms checked before agents make decision to consume. Can be set between "1" and the total number of firms.\\
    Consumption satisfaction & \multicolumn{1}{c}{0.01} & \multicolumn{1}{c}{[0, 1)} & Used to measure satisfaction gained with consumption. \\
    \textbf{Families} & \multicolumn{1}{c}{} & \multicolumn{1}{c}{} &  \\
    Real estate market & \multicolumn{1}{c}{0.021} & \multicolumn{1}{c}{(0, 1)} &  Percentage of families' entering real estate market\\
    \textbf{Government } & \multicolumn{1}{c}{} & \multicolumn{1}{c}{} &  \\
    Consumption tax & \multicolumn{1}{c}{0.21} & \multicolumn{1}{c}{(0, 1)} & Tax on consumption. \\
    \bottomrule
    \end{tabular}%
  \label{tab:tab2}%
\end{table*}%

The proposed model contains a very small number of exogenous parameters. The parameters help understand how the model mechanisms influence its results. Parameter $\alpha$, for example, can be set to 1 so that its effect is zero. By reducing the parameter successively by 0.1, one can observe the effect of increasingly less productive workers. The same understanding of relevance can be made with the $\beta$ parameter, or the rate of consumption tax. This construction offers some flexibility to the modeler. This flexibility is most relevant if the goal is to increase understanding of the problem that is modeled and the model is used as a guiding tool for decision-making, or as a methodology to discuss 'what if' questions.

\subsection{Limitations}

The limitation of this study arises from the difficulty of finding complete, integrated, and simple models that could be used as initial steps to be expanded and adapted by following researchers. In fact, despite the models of Lengnick \cite{lengnick_agent-based_2013} and Gaffeo et al. \cite{gaffeo_adaptive_2008}, all others are specific to a single market, such as energy \cite{koesrindartoto_agent-based_2005}, finance \cite{feng_linking_2012} or labor market \cite{seppecher_flexibility_2012}; or are too complex \cite{dawid_agent-based_2014,van_der_hoog_production_2008}.

Thus, the task of investigating a specific phenomenon (in this case, the spatial influence of consumption taxes), under a single integrated model, requires that all processes (firm production firm, goods and labor market) and all its associated parameters are explained, theoretically sound and numerically compatible. In practice, this requires that the modeler get results that are consistent with the literature for indicators such as inflation, GDP growth, unemployment, household income, simultaneously and  temporally consistent.

The question put to modelers is of an epistemological nature. How can we determine which are the central elements of the phenomenon, which must be present, and what are the accessory elements? At what point, simplifying the process can take place and where there is significant change of the observed phenomenon?

Besides this general limitation, this version of the model also does not include the credit market, demographic changes nor investment in social capital.

\section{Results and discussion}\label{sec3}

\begin{figure}[H]
	\centering
	\includegraphics[width=9cm]{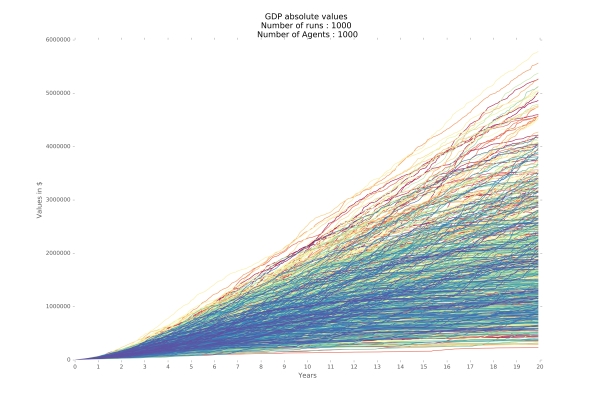}
	\caption{GDP growth results, 1,000 iterations, for one region.}
	\label{fig2}
\end{figure}

\begin{figure}[H]
	\centering
	\includegraphics[width=9cm]{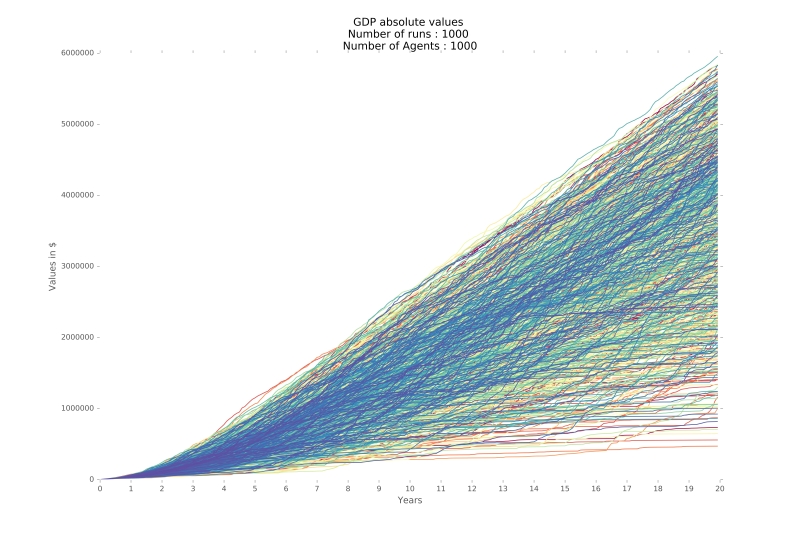}
	\caption{GDP growth results, 1,000 iterations, for seven regions.}
	\label{fig3}
\end{figure}

\begin{figure}[!t]
	\centering
	\includegraphics[width=9cm]{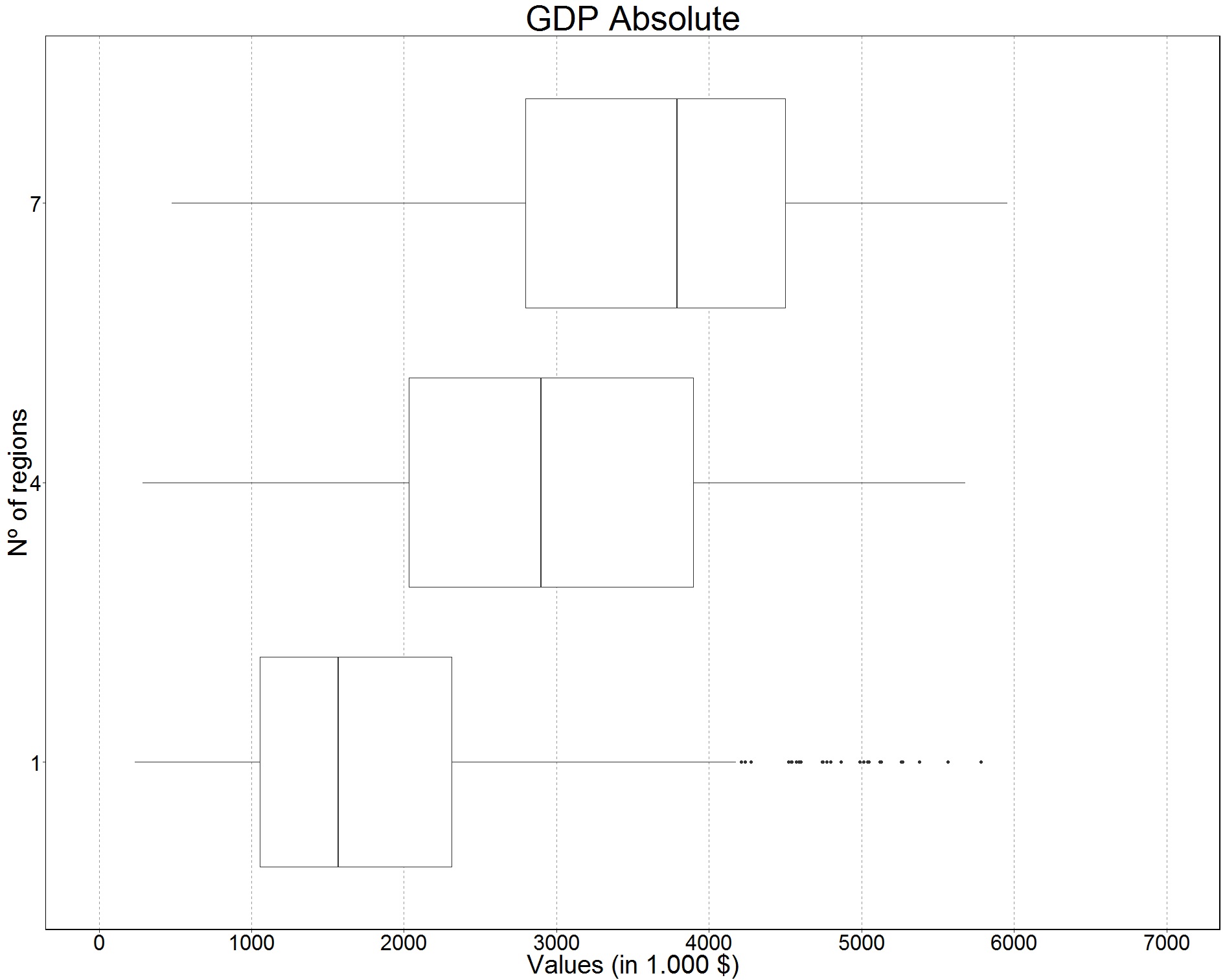}
	\caption{BoxPlot of GDP for the last month of the simulation, 1,000 iterations, for the three regions.}
	\label{fig4}
\end{figure}

The most thriving economy is the model with seven regions (\autoref{fig3}), on average with median GDP only 30\% higher when compared to the model with four regions. This model with four regions, in turn, achieves results that are 38\% above the model with a single region (\autoref{fig2}). The variability is higher for the model with seven regions, vis-\`a-vis the one region model (\autoref{tab3} and  \autoref{fig4}).

\begin{table}[htbp]
	\centering
    \caption{Median, first and third quartiles for the last month of the simulations (1,000 iterations) of Gini coefficient, GDP, QLI and families' wealth for each regional design}
    \begin{tabular}{rrrr}
    \toprule
    \textit{\textbf{}} & \textit{\textbf{Gini}} & \textit{\textbf{}} & \textit{\textbf{}} \\
    \midrule
    \multicolumn{1}{c}{Regions} & \multicolumn{1}{c}{0.25} & \multicolumn{1}{c}{Median} & \multicolumn{1}{c}{0.75} \\
    \hline
    \multicolumn{1}{c}{1} & \multicolumn{1}{c}{0.890} & \multicolumn{1}{c}{0.916} & \multicolumn{1}{c}{0.932} \\
    \multicolumn{1}{c}{4} & \multicolumn{1}{c}{0.925} & \multicolumn{1}{c}{0.939} & \multicolumn{1}{c}{0.946} \\
    \multicolumn{1}{c}{7} & \multicolumn{1}{c}{0.935} & \multicolumn{1}{c}{0.944} & \multicolumn{1}{c}{0.950} \\
    \hline
    \textit{\textbf{}} & \textit{\textbf{GDP}} & \textit{\textbf{}} & \textit{\textbf{}} \\
    \hline
    \multicolumn{1}{c}{Regions} & \multicolumn{1}{c}{0.25} & \multicolumn{1}{c}{Median} & \multicolumn{1}{c}{0.75} \\
    \hline
    \multicolumn{1}{c}{1} & \multicolumn{1}{c}{1,056,341} & \multicolumn{1}{c}{1,568,746} & \multicolumn{1}{c}{2,314,751} \\
    \multicolumn{1}{c}{4} & \multicolumn{1}{c}{2,029,562} & \multicolumn{1}{c}{2,904,486} & \multicolumn{1}{c}{3,897,903} \\
    \multicolumn{1}{c}{7} & \multicolumn{1}{c}{2,794,786} & \multicolumn{1}{c}{3,788,903} & \multicolumn{1}{c}{4,501,469} \\
    \hline
    \textit{\textbf{}} & \textit{\textbf{QLI}} & \textit{\textbf{}} & \textit{\textbf{}} \\
    \hline
    \multicolumn{1}{c}{Regions} & \multicolumn{1}{c}{0.25} & \multicolumn{1}{c}{Median} & \multicolumn{1}{c}{0.75} \\
    \hline
    \multicolumn{1}{c}{1} & \multicolumn{1}{c}{223.2} & \multicolumn{1}{c}{331.8} & \multicolumn{1}{c}{487.3} \\
    \multicolumn{1}{c}{4} & \multicolumn{1}{c}{425.0} & \multicolumn{1}{c}{608.9} & \multicolumn{1}{c}{820.9} \\
    \multicolumn{1}{c}{7} & \multicolumn{1}{c}{562.1} & \multicolumn{1}{c}{761.8} & \multicolumn{1}{c}{945.9} \\
    \hline
    \textit{\textbf{}} & \textit{\textbf{}} & \textit{\textbf{Families' wealth}} & \textit{\textbf{}} \\
    \hline
    \multicolumn{1}{c}{Regions} & \multicolumn{1}{c}{0.25} & \multicolumn{1}{c}{Median} & \multicolumn{1}{c}{0.75} \\
    \hline
    \multicolumn{1}{c}{1} & \multicolumn{1}{c}{589} & \multicolumn{1}{c}{19,757} & \multicolumn{1}{c}{115,815} \\
    \multicolumn{1}{c}{4} & \multicolumn{1}{c}{32,126} & \multicolumn{1}{c}{253,450} & \multicolumn{1}{c}{893,763} \\
    \multicolumn{1}{c}{7} & \multicolumn{1}{c}{82,190} & \multicolumn{1}{c}{573,573} & \multicolumn{1}{c}{1,645,296} \\
    \bottomrule
    \end{tabular}%
  \label{tab3}%
\end{table}%

\begin{figure}[!t]
\centering
\includegraphics[width=9cm]{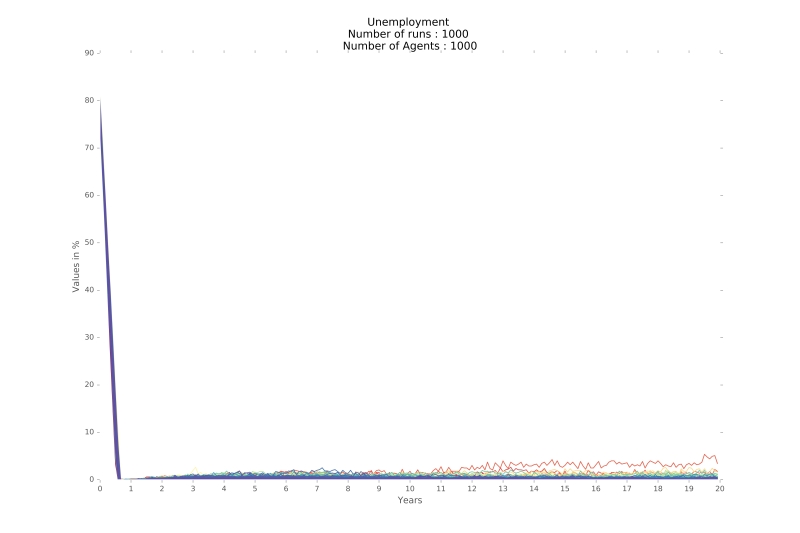}
\caption{Unemployment, 1,000 iterations and one region.}
\label{fig5}
\end{figure}

\begin{figure}[!t]
	\centering
	\includegraphics[width=9cm]{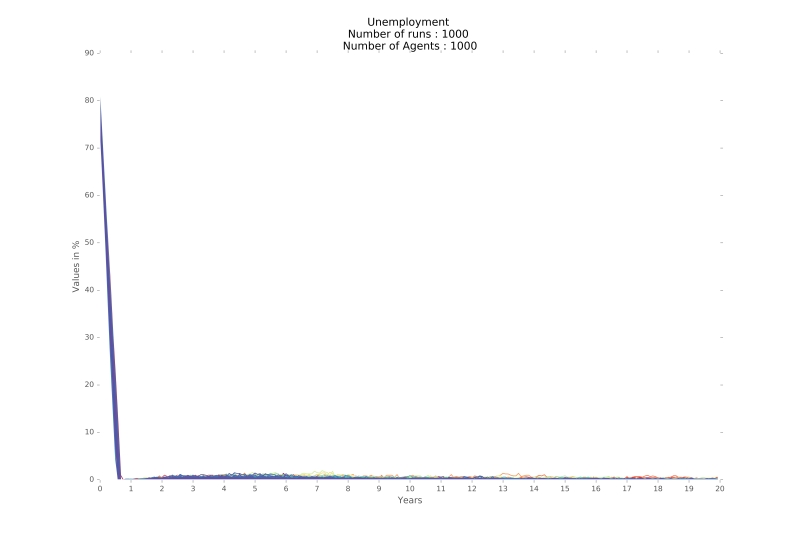}
	\caption{Unemployment, 1,000 iterations and seven regions}
	\label{fig6}
\end{figure}

Considering the labor market, the economies converge towards full employment, keeping a cycle of very low unemployment, throughout the period, for the three regional designs show similar results (\autoref{fig5} and \autoref{fig6}).

\begin{figure}[!t]
\centering
\includegraphics[width=9cm]{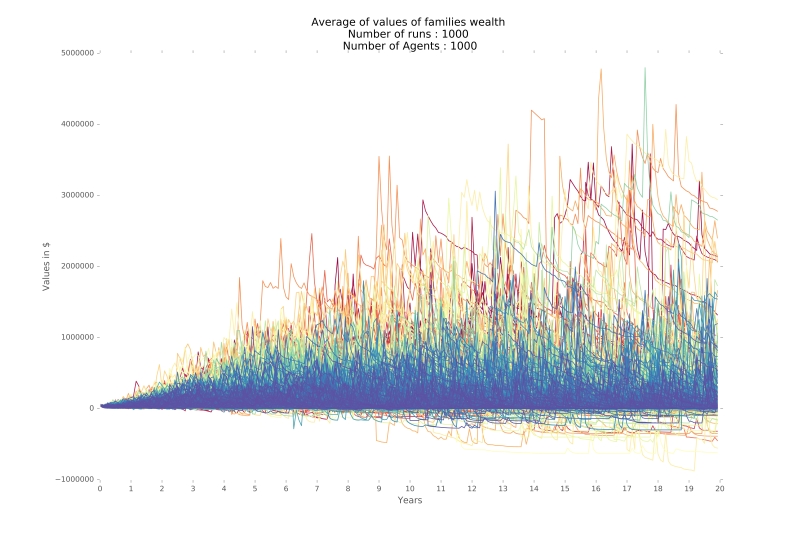}
\caption{Families' wealth, 1,000 iterations, for one region.}
\label{fig7}
\end{figure}

\begin{figure}[!t]
	\centering
	\includegraphics[width=9cm]{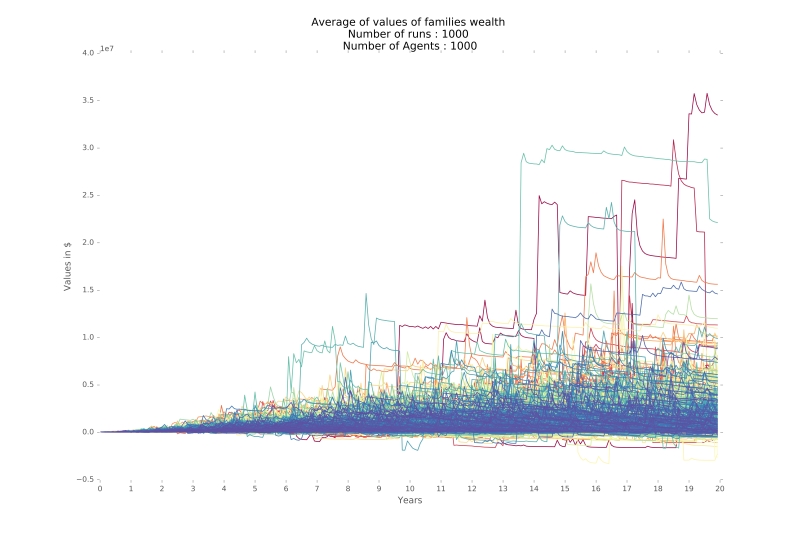}
	\caption{Families' wealth, 1,000 iterations, for seven regions.}
	\label{fig8}
\end{figure}

\begin{figure}[!t]
\centering
\includegraphics[width=9cm]{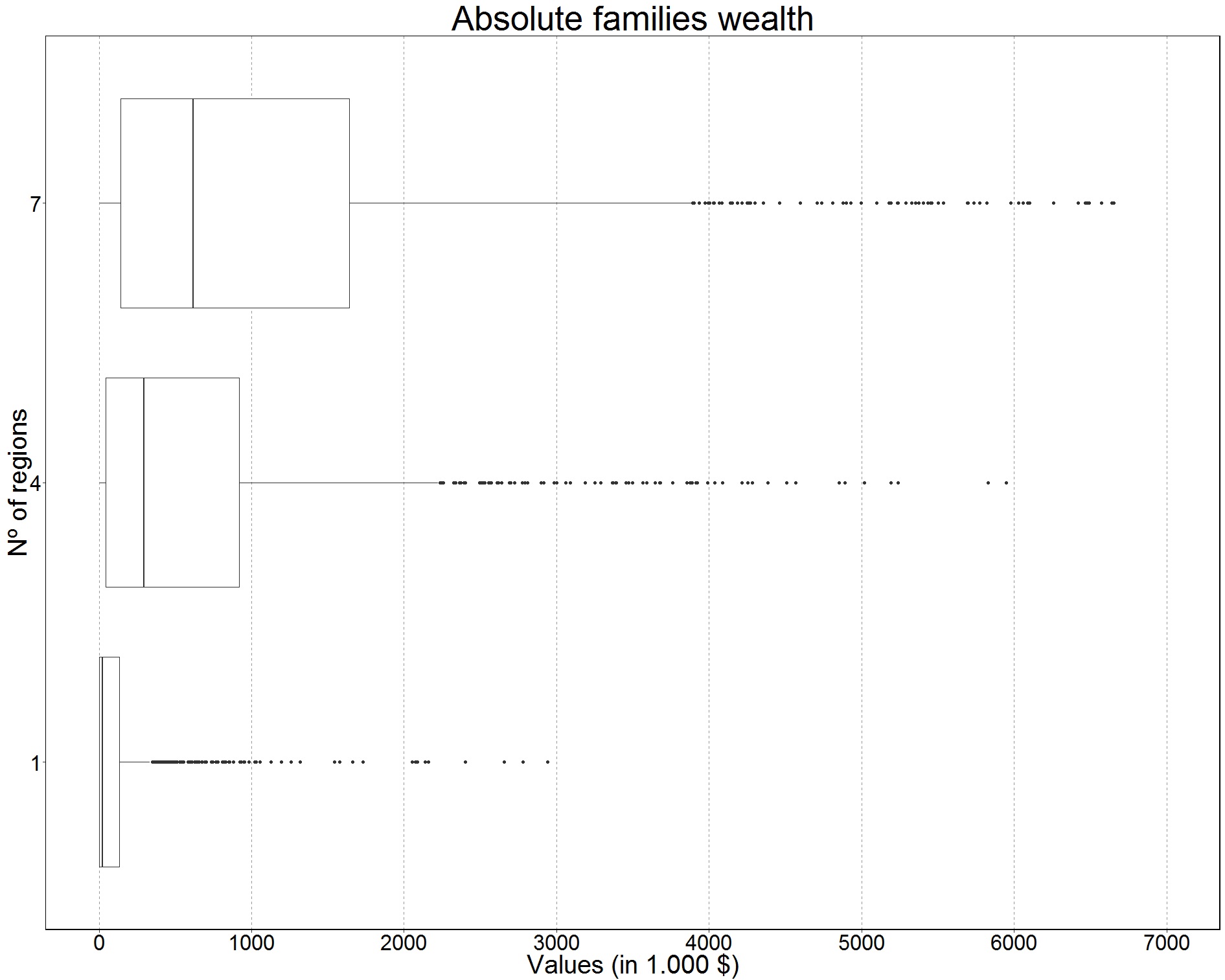}
\caption{BoxPlot of families' wealth for the last month of the simulations, 1,000 iterations, for the three regional designs}
\label{fig9}
\end{figure}

Household income varies significantly among the three regional designs, for one region (\autoref{fig7}) variation is of lower magnitude when compared to the model with seven regions (\autoref{fig8}) through 1.000 iterations. The median, first and third quartiles are higher for the model with seven regions, vis-\`a-vis the model with only one region (\autoref{tab3} and \autoref{fig6}). Although the dispersion is higher for the model with seven regions, the value of the first quartile is slightly less than the median with a single region (\autoref{fig9}).

\begin{figure}[!t]
\centering
\includegraphics[width=9cm]{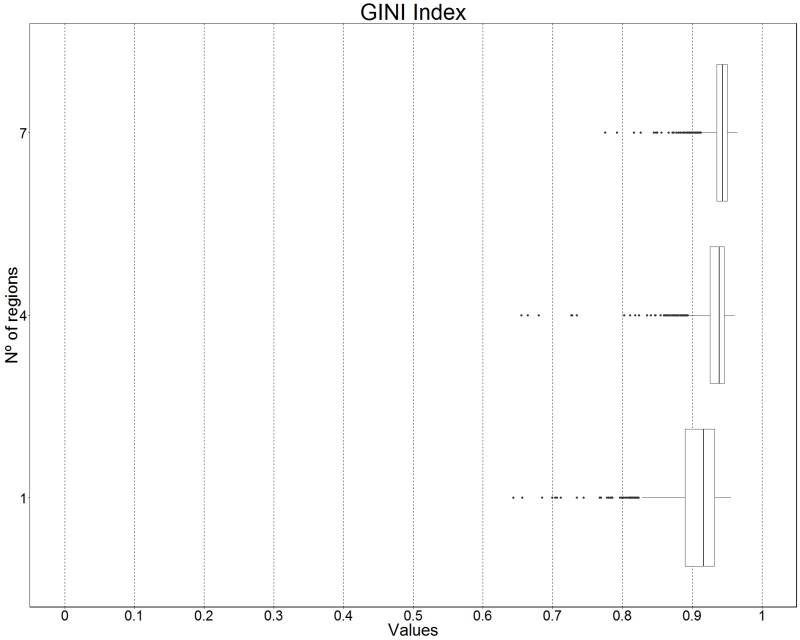}
\caption{BoxPlot of the Gini coefficient for the last month of the simulation, 1,000 iterations, for the three regional design.}
\label{fig10}
\end{figure}

\begin{figure}[!t]
\centering
\includegraphics[width=9cm]{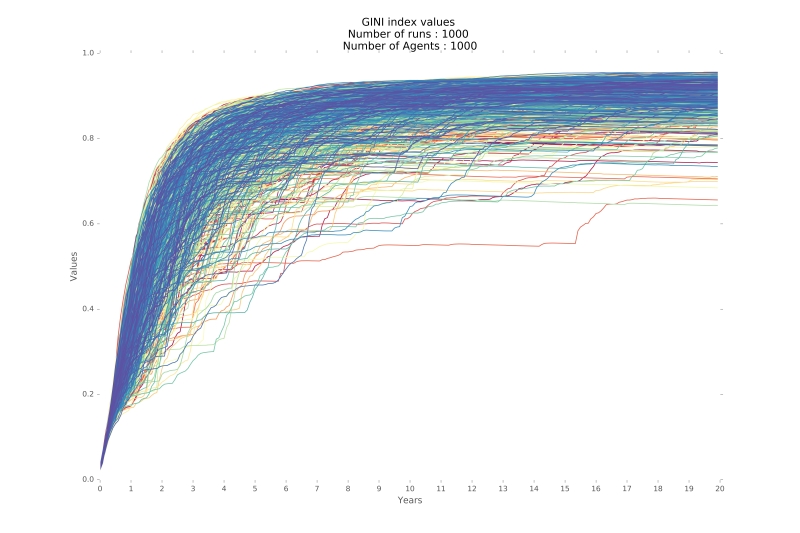}
\caption{Results for the Gini coefficient, 1,000 iterations, one region.}
\label{fig11}
\end{figure}

\begin{figure}[!t]
	\centering
	\includegraphics[width=9cm]{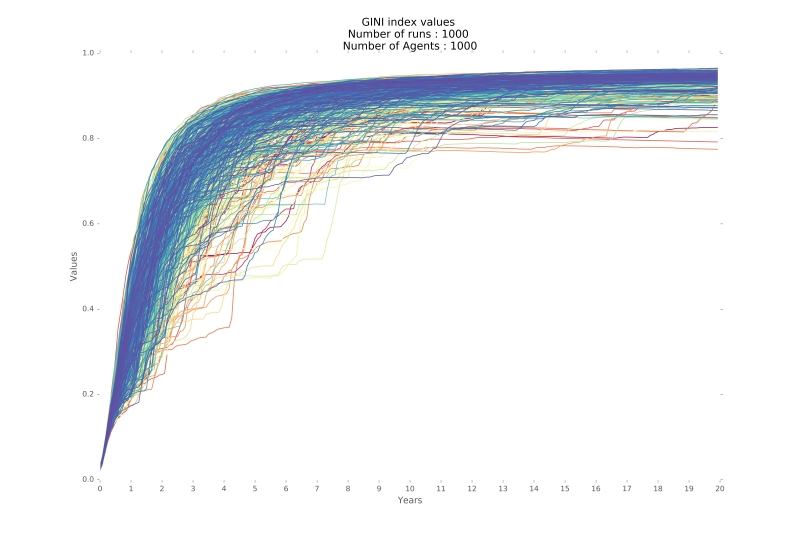}
	\caption{Results for the Gini coefficient, 1,000 iterations, seven regions.}
	\label{fig12}
\end{figure}

The Gini coefficient is computed on the utility of the families. Utility is directly proportional to the cumulative consumption of the families. The GINI coefficient reaches a higher level in the model with 7 regions when compared to the two other models (\autoref{fig10}). Moreover, the behavior of the coefficient in the 1,000 iterations has a similar pattern of variability (\autoref{fig11} and \autoref{fig12}) with slightly higher standard deviation. 

\begin{figure}[!t]
\centering
\includegraphics[width=9cm]{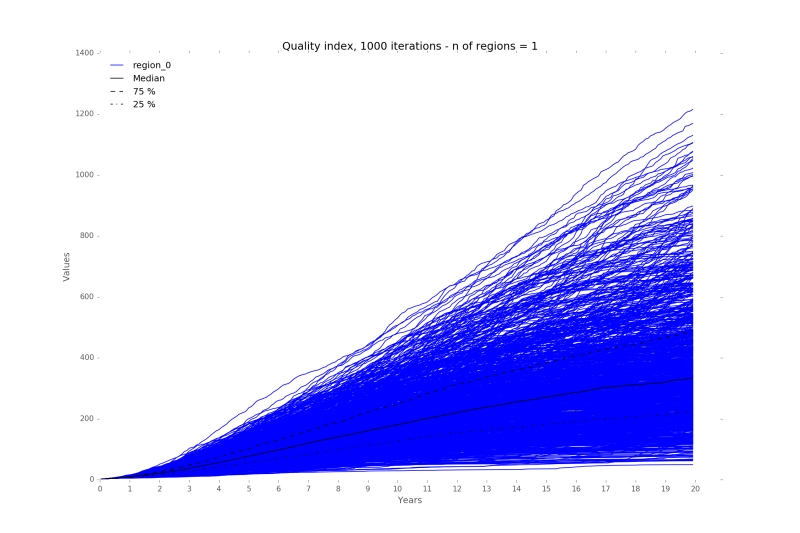}
\caption{Results of the Quality of Life Index (QLI), 1,000 iterations, for one region.}
\label{fig13}
\end{figure}

\begin{figure}[!t]
	\centering
	\includegraphics[width=9cm]{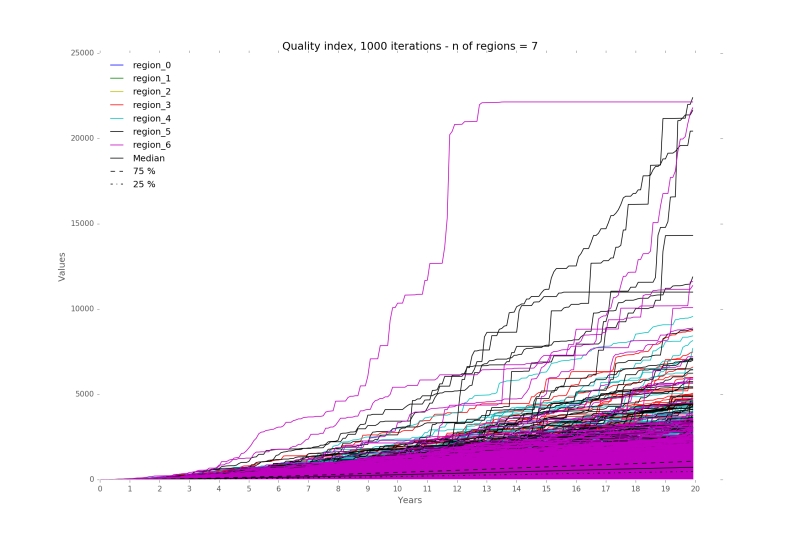}
	\caption{Results of the Quality of Life Index (QLI), 1,000 iterations, for seven regions.}
	\label{fig14}
\end{figure}

\begin{figure}[!t]
\centering
\includegraphics[width=9cm]{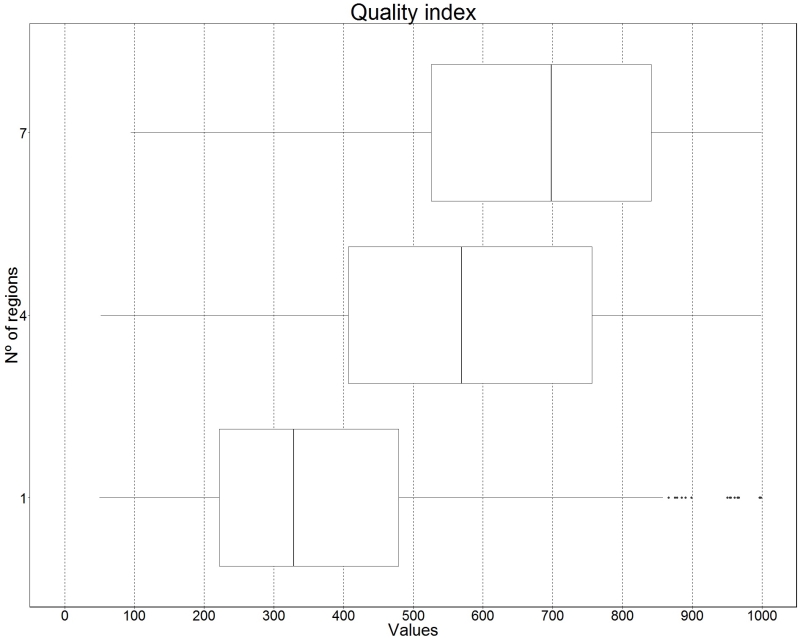}
\caption{BoxPlot of the QLI, for the last month of the simulation, 1,000 iterations for the three regional designs.}
\label{fig15}
\end{figure}

Finally, the basic indicator to compare the performance of the models is the Quality of Life Index for each simulated design for one region (\autoref{fig13}) and for seven regions (\autoref{fig14}). The median and the quartiles are higher for the model with seven regions  vis-\`a-vis the other models (\autoref{fig15}).

The model results indicate that changes in administrative boundaries have led to robust changes among the three considered region design. According to the procedures described, the dynamism of the real estate market, namely, household mobility in the simulations with more than one region, was relevant to the results.

In the absence of a credit market, families with income level below the median become sellers in the real estate market. Thus, these families capitalize on the sale of homes whose prices increased along with the quality of life in the region, and migrate toward regions with poorer quality. This movement is partly counterbalanced by families trying to migrate in search of better quality.

As a result, the models with subdivisions lead to regions that are less populated and have better quality of life and at the same time, more populated regions that have worse QLI (\autoref{tab4}).

\begin{table}[htbp]
  \centering
  \caption{Median and standard deviation of the maximum and minimum regional values for each simulation for QLI and population, 1,000 iterations, by regional design.}
    \begin{tabular}{p{0.9cm}p{0.7cm}p{1.2cm}p{1cm}p{1.2cm}p{0.9cm}}
    \toprule
    \multicolumn{1}{c}{\textbf{Regions}} & \multicolumn{1}{c}{\textbf{}} & \textbf{QLI median} & \textbf{QLI std} & \textbf{Pop. Median} & \multicolumn{1}{c}{\textbf{Pop. Std}} \\
    \midrule
    \multicolumn{1}{l}{1} & Max & 333.5 & 210.9 & 1,000 &  \\
    \multicolumn{1}{l}{} & Min & 333.5 & 210.9 & 1,000 &  \\
    \multicolumn{1}{l}{4} & Max & 860.2 & 430.2 & 207 & \multicolumn{1}{c}{30.3} \\
    \multicolumn{1}{l}{} & Min & 423.2 & 198 & 288 & \multicolumn{1}{c}{29.2} \\
    \multicolumn{1}{l}{7} & Max & 1,499.2 & 2,047.7 & 40 & \multicolumn{1}{c}{70.2} \\
    \multicolumn{1}{l}{} & Min & 343.9 & 195.3 & 83 & \multicolumn{1}{c}{89.9} \\
    \bottomrule
    \end{tabular}%
  \label{tab4}%
\end{table}%

Finally, it is noteworthy the trade off between the results for the three models. While the model with seven regions is more dynamic, more productive and wealthier, it is also more heterogeneous. The model with one region, in turn, is more harmonic but less vigorous.

The underlying assumption of the authors – that the model with one region would be more efficient from the standpoint of conurbation regions – was not observed with the present configuration. Especially given the strength and mobility of the real estate market that concentrates smaller populations in regions with higher quality and larger populations in areas with poorer quality. However, the research question – that is, if administrative boundaries influence the economic and fiscal dynamic of the regions – can be confirmed.

Yet, the results indicate the wealth of possibilities of analysis of the economic system from heterogeneous agents and firms in an environment that is continually changing.

Finally, given the process of creating artificial economies, at each loop iteration the agents and firms are completely different. Thus, the next phase of research, which is the model application to real data, will input actual data as attributes of the economy and thus reduce the variability of results.

\subsection{Sensitivity analysis and robustness}

The sensitivity analysis is central in building simulation models to ensure that the model is structurally consistent and does not depend solely on a particular parameter, which is adjusted for a specific value. Furthermore, the sensitivity analysis may serve as an analytical tool to show how and with which magnitude certain configurations and model processes change trends and results.

The sensitivity analysis made was based on the variation of the model parameters in 10 different values between their minimum and maximum values (\autoref{tab5}). As random numbers influence the model results, comparing the results of different iterations (model runs) is only possible if we use the same seed. Thus, if the model is run several times with the same parameter and the same seed, the same results will be produced. Therefore, when the modeler changes the parameters, variations in sensitivity analysis results will be a result of the model structure and not of the random number generator.\footnote{The results of the sensitivity analysis were obtained using a fixed seed.} 

The change of the parameters was performed separately (one parameter at a time) with the other parameters maintained at their default values, defined in a first exploratory analysis.

\begin{table*}[htbp]
  \centering
  \caption{Simulation parameter values used in the sensitivity analysis}

    \begin{tabular}{rrrrrrrrrrr}
    \toprule
    \textbf{Parameters} & \textbf{Values } & \textbf{} & \textbf{} & \textbf{} & \textbf{} & \textbf{} & \textbf{} & \textbf{} & \textbf{} & \textbf{} \\
    \midrule
    \multicolumn{1}{l}{Alpha} & \multicolumn{1}{c}{0.1} & \multicolumn{1}{c}{0.14} & \multicolumn{1}{c}{0.19} & \multicolumn{1}{c}{0.23} & \multicolumn{1}{c}{0.28} & \multicolumn{1}{c}{0.32} & \multicolumn{1}{c}{0.37} & \multicolumn{1}{c}{0.41} & \multicolumn{1}{c}{0.46} & \multicolumn{1}{c}{0.5} \\
    \multicolumn{1}{l}{Beta} & \multicolumn{1}{c}{0.5} & \multicolumn{1}{c}{0.55} & \multicolumn{1}{c}{0.61} & \multicolumn{1}{c}{0.66} & \multicolumn{1}{c}{0.72} & \multicolumn{1}{c}{0.77} & \multicolumn{1}{c}{0.83} & \multicolumn{1}{c}{0.88} & \multicolumn{1}{c}{0.94} & \multicolumn{1}{c}{0.99} \\
    \multicolumn{1}{l}{Quantity to change prices} & \multicolumn{1}{c}{10} & \multicolumn{1}{c}{42} & \multicolumn{1}{c}{74} & \multicolumn{1}{c}{107} & \multicolumn{1}{c}{139} & \multicolumn{1}{c}{171} & \multicolumn{1}{c}{203} & \multicolumn{1}{c}{235} & \multicolumn{1}{c}{268} & \multicolumn{1}{c}{300} \\
    \multicolumn{1}{l}{\textit{Mark-up}} & \multicolumn{1}{c}{0.01} & \multicolumn{1}{c}{0.04} & \multicolumn{1}{c}{0.06} & \multicolumn{1}{c}{0.09} & \multicolumn{1}{c}{0.12} & \multicolumn{1}{c}{0.14} & \multicolumn{1}{c}{0.17} & \multicolumn{1}{c}{0.2} & \multicolumn{1}{c}{0.22} & \multicolumn{1}{c}{0.25} \\
    \multicolumn{1}{l}{Labor market entry} & \multicolumn{1}{c}{0.1} & \multicolumn{1}{c}{0.14} & \multicolumn{1}{c}{0.19} & \multicolumn{1}{c}{0.23} & \multicolumn{1}{c}{0.28} & \multicolumn{1}{c}{0.32} & \multicolumn{1}{c}{0.37} & \multicolumn{1}{c}{0.41} & \multicolumn{1}{c}{0.46} & \multicolumn{1}{c}{0.5} \\
    \multicolumn{1}{l}{Market size} & \multicolumn{1}{c}{1} & \multicolumn{1}{c}{3} & \multicolumn{1}{c}{5} & \multicolumn{1}{c}{7} & \multicolumn{1}{c}{10} & \multicolumn{1}{c}{15} & \multicolumn{1}{c}{30} & \multicolumn{1}{c}{50} & \multicolumn{1}{c}{70} & \multicolumn{1}{c}{110} \\
    \multicolumn{1}{l}{Real estate entry} & \multicolumn{1}{c}{0.01} & \multicolumn{1}{c}{0.02} & \multicolumn{1}{c}{0.03} & \multicolumn{1}{c}{0.04} & \multicolumn{1}{c}{0.05} & \multicolumn{1}{c}{0.06} & \multicolumn{1}{c}{0.07} & \multicolumn{1}{c}{0.08} & \multicolumn{1}{c}{0.09} & \multicolumn{1}{c}{0.1} \\
    \multicolumn{1}{l}{Tax consumption} & \multicolumn{1}{c}{0.01} & \multicolumn{1}{c}{0.06} & \multicolumn{1}{c}{0.11} & \multicolumn{1}{c}{0.16} & \multicolumn{1}{c}{0.21} & \multicolumn{1}{c}{0.25} & \multicolumn{1}{c}{0.3} & \multicolumn{1}{c}{0.35} & \multicolumn{1}{c}{0.4} & \multicolumn{1}{c}{0.45} \\
    \bottomrule
    \end{tabular}%
  \label{tab5}%
\end{table*}%

Given that the premise is to create a model (or simulation machine), the sensitivity analysis furthers our understanding of the model. Of course, we also varied the number of regions (1, 4, 7). 

\subsubsection{Alpha}

The variation of each of the parameters affects differently the results of the simulations. The alpha parameter – which evaluates worker productivity – for example, leads to higher values of total GDP when values are between 0.32 and 0.37. 

\begin{figure}[!t]
\centering
\includegraphics[width=9cm]{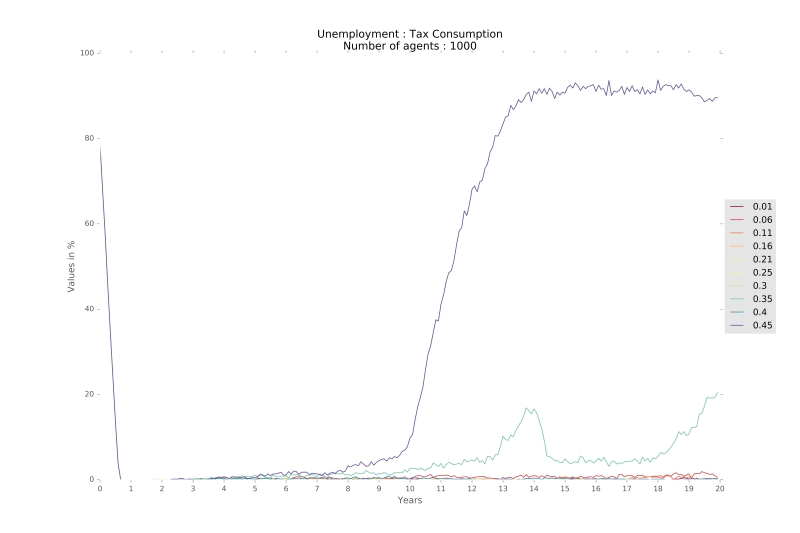}
\caption{Results of alpha variation on unemployment, one region.}
\label{fig16}
\end{figure}

\begin{figure}[!t]
	\centering
	\includegraphics[width=9cm]{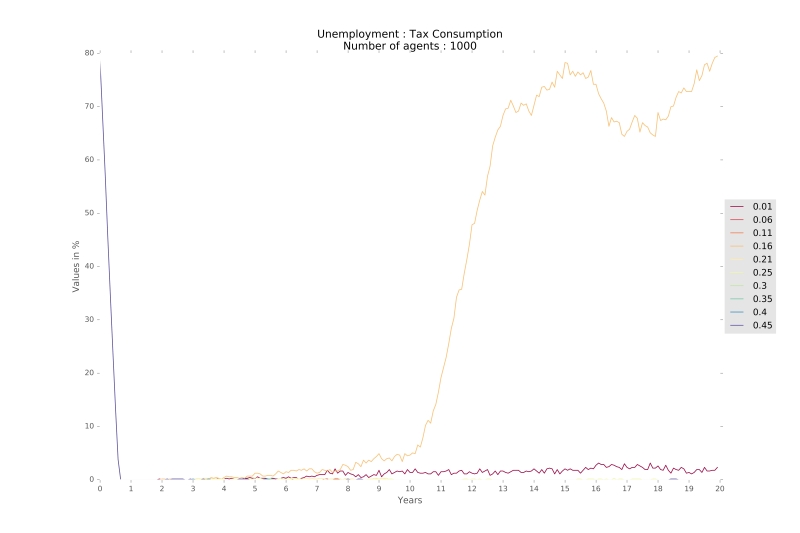}
	\caption{Results of alpha variation on unemployment, seven regions.}
	\label{fig17}
\end{figure}

Considering unemployment, the alpha parameter provides full employment, conditioned to other parameter's default value. Figures \autoref{fig16} and \autoref{fig17} shows that unemployment converges quickly towards full employment. The behavior was similar for all regions. The variation in productivity and unemployment indicate that when worker productivity is very high, there is an excess supply that is not absorbed by the market.

\begin{figure}[!t]
\centering
\includegraphics[width=9cm]{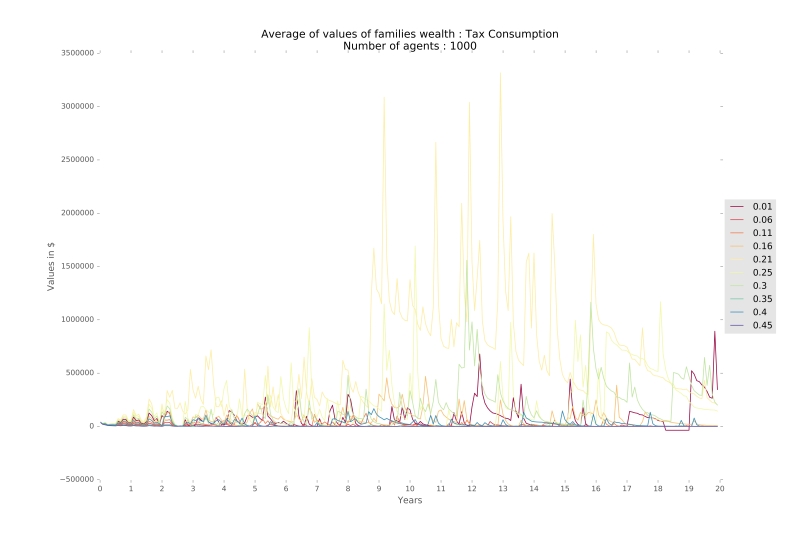}
\caption{Firms' profit variation results for Alpha 0.1 and 0.9, and one region}
\label{fig18}
\end{figure}

There is greater variation in firms' profits when alpha is smaller and more stability when alpha is higher (\autoref{fig18}). 

The value of alpha at 0.35 results in a good balance of resources in the economy among firms and households, on average, for all the region designs.

The Gini coefficient is slightly higher for higher values of alpha. Finally, prices rise slightly to higher values of alpha.

\subsubsection{Beta}

\begin{figure}[!t]
\centering
\includegraphics[width=9cm]{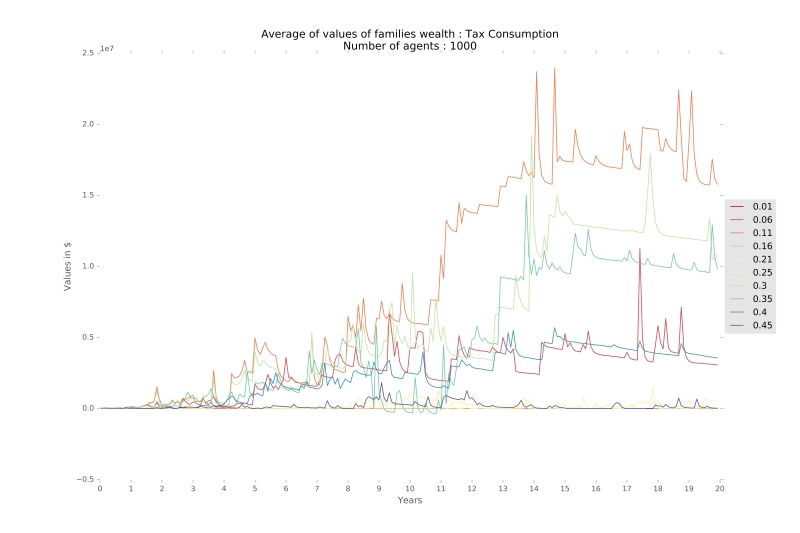}
\caption{Results for the variation of parameter beta on unemployment values for one region.}
\label{fig19}
\end{figure}

\begin{figure}[!t]
\centering
\includegraphics[width=9cm]{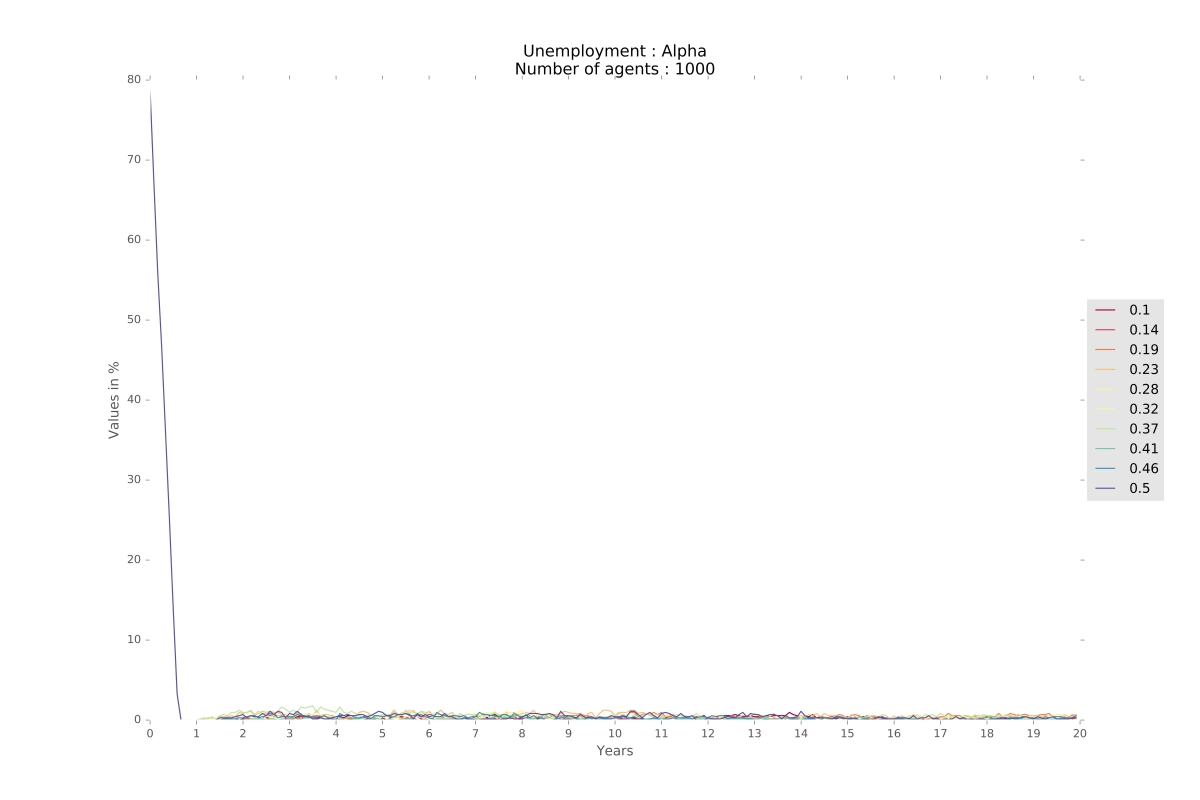}
\caption{Results for the variation of parameter beta on unemployment values for seven regions.}
\label{fig20}
\end{figure}

The beta parameter – which controls the propensity to consume of households – influences strongly the economy. In fact, higher values of beta (lower discount at maximum limit of household spending) or lower levels of beta, which restricts consumption, lead to high and persistent levels of unemployment (\autoref{fig19} and \autoref{fig20}), especially in the model with one region, where the dynamics is more dependent on the goods market. Low values also keep firms' profits close to zero.

\begin{figure}[!t]
	\centering
	\includegraphics[width=9cm]{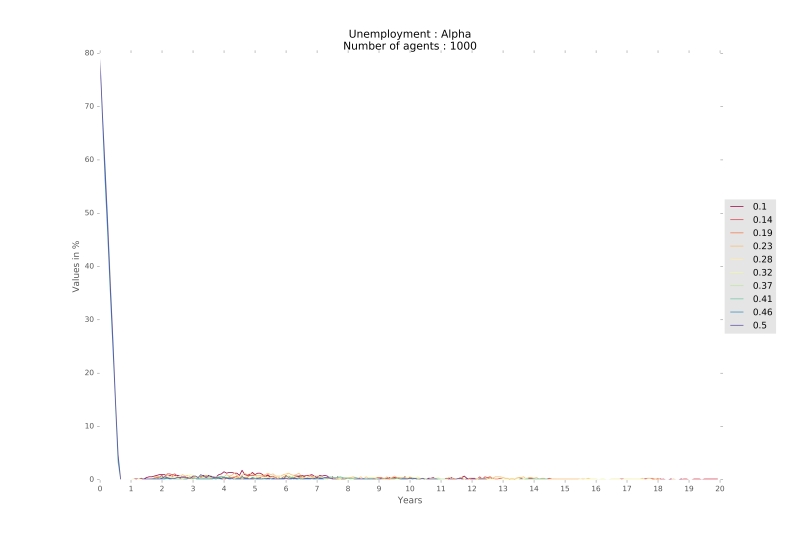}
	\caption{Variation in the results of the Gini coefficient for beta values 0.1 and 0.9, one region.}
	\label{fig21}
\end{figure}

The impact of beta in the Gini coefficient is relevant and similar among the regional designs used. For low values of beta and low household consumption, Gini rises gradually, reaching a maximum at about 0.5. However, when beta has a value of 0.99, inequality rises steeply to reach values close to 0.90 at the end of the period (\autoref{fig21}). 

\subsubsection{Tax on consumption}

\begin{figure}[!t]
	\centering
	\includegraphics[width=9cm]{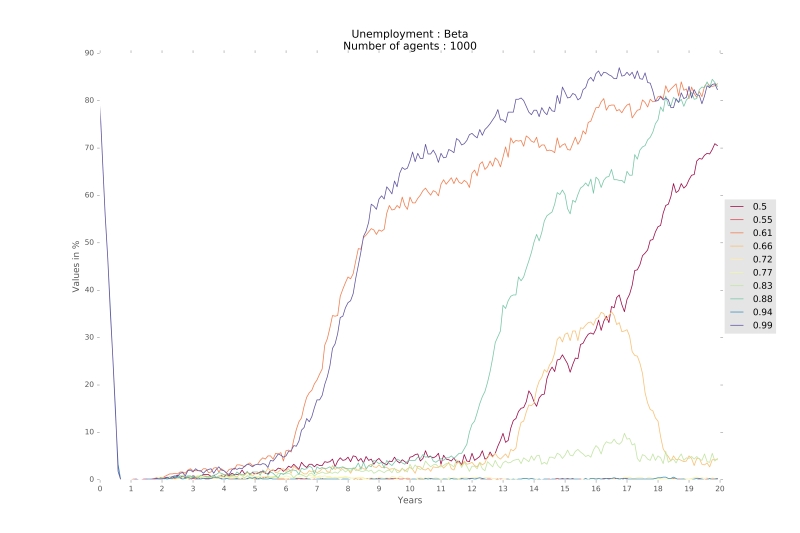}
	\caption{Results of unemployment for various consumption tax rate values, one region.}
	\label{fig22}
\end{figure}

\begin{figure}[!t]
	\centering
	\includegraphics[width=9cm]{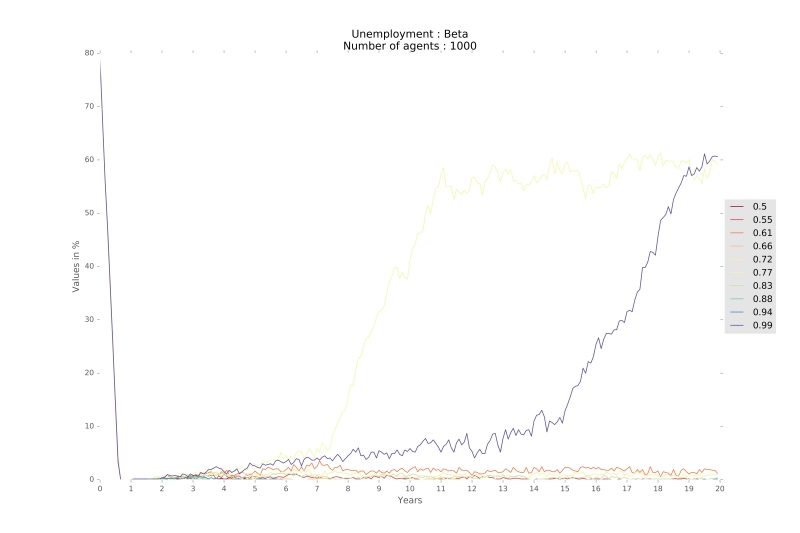}
	\caption{Results of unemployment for various consumption tax rate values, seven regions.}
	\label{fig23}
\end{figure}

\begin{figure}[!t]
	\centering
	\includegraphics[width=9cm]{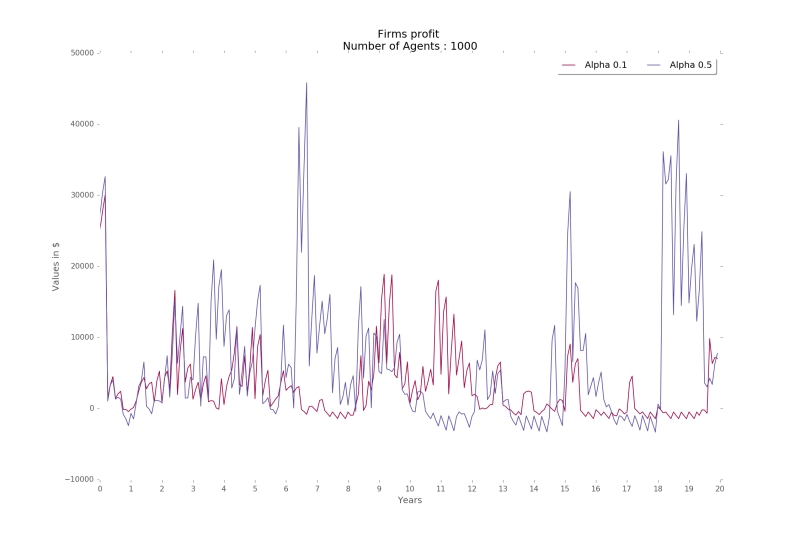}
	\caption{Families' wealth results for various consumption tax rate, one region.}
	\label{fig24}
\end{figure}

\begin{figure}[!t]
	\centering
	\includegraphics[width=9cm]{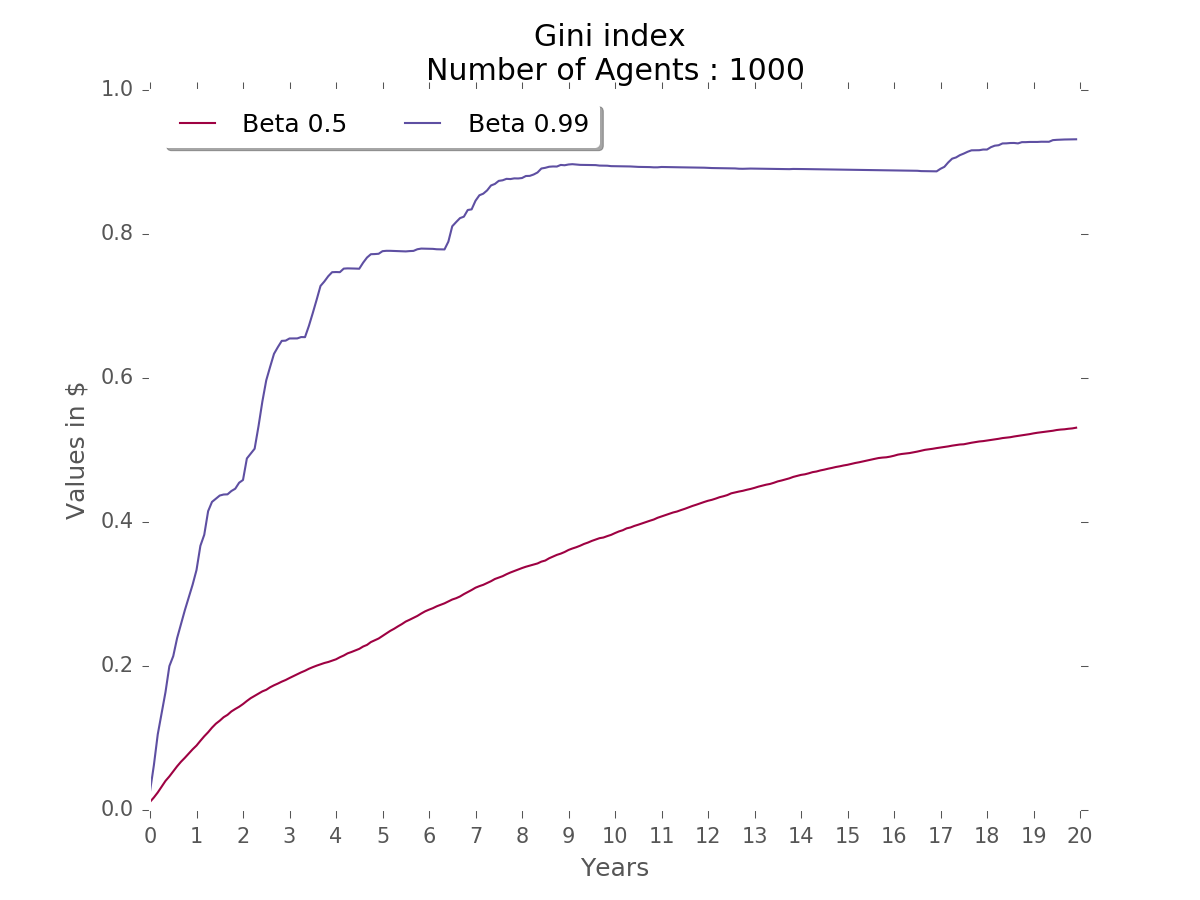}
	\caption{Families' wealth results for various consumption tax rate, seven regions.}
	\label{fig25}
\end{figure}

As expected, the value of the tax rate influences the economy on many levels. Lower rates lead to lower unemployment, but influence little when below 0.5 (\autoref{fig22} and \autoref{fig23}). Very high tax rates bring hyperinflation, widespread unemployment and a significant drop in revenues and profits of firms. However, given that the amounts collected in taxes are applied in the same regions where collected, QLI improves, and consequently, property prices increase accordingly. Household income (\autoref{fig24} and \autoref{fig25}) and GDP are higher for intermediate values of tax rate

\subsubsection{Other parameters}

A sensitivity analysis was performed for each exogenous parameters of the model, with lower relative impact compared to parameters alpha, beta and the tax rate.

The level of the inventory that triggers changes in prices, for example, seems to impact slightly on the evolution of the price index, delaying its increase.

The frequency with which firms enter the labor market affects the speed of adjustment in the labor market. When parameter values are higher – taking longer to enter the labor market – unemployment is only insignificant at the end of the period. When the entry of firms is frequent, full employment is achieved within months.

The change in the mark-up value, i.e., the percentage increase in product prices of firms when their stock is low – does not greatly change the profit levels of the firms. However, very high mark-up rates, lead to uncontrolled inflation after some time.

The size of the market checked by consumers when they go shopping does not interfere in the results, with little loss of momentum when size is restricted to only one firm.

Finally, the percentage of families entering the housing market seems to have little influence. When all families are on the market all the time, there is a small reduction in household income.

Thus, we understand that the variation of the results of the model given by the variation of the parameters is in line with the underlying theory. In addition, there is no change of parameters that cause different or unexpected behavior of the model. Thus, we believe it indicates the robustness of the model, as described.

\section{The theoretical-methodological approach: possibilities for future research}\label{sec4}

This section describes various additions to the model that could be implemented with relatively small and simple changes in the current code. Given the seminal methodological trait of this paper, we thought the model should be developed in its simplest form possible, following the KISS logic ('keep it simple, stupid'). Eventually, it could evolve into the KIDS form ('keep it descriptive, stupid'), formulated by Edmonds and Moss \cite{edmonds_kiss_2005}.

The immediate interest of the authors, it is to apply it for the Federal District region, in Brazil. Empirical data would be used in the initial configuration of the model, namely: actual municipal boundaries, specific spatially-bound demographic patterns, actual companies attributes and location, and supply of skilled labor. The following step would be to validate the empirical model for a given time line, seeking correlation or similarity between the evolution of observed indicators and those produced by the model. Finally, after validation, the model could be effectively used to implement public policy alternatives.

The actual realm of research possibilities are detailed below, following the KIDS argument of Edmonds and Moss \cite{edmonds_kiss_2005}.
\begin{enumerate}
\item Implement demographic change, with processes that describe birth, deaths and families’ creation in order to become a more dynamic and real model while enabling results for specific demographic cohorts. In addition, inter-temporal analyzes involving inheritance (of wealth or social capital), could also be tested;
\item Another relatively simple alternative is the inclusion of updating workers qualification (years of study), deducing investment from their resources;
\item The credit market, with production and consumer financing possibility is also relevant to make the model closer to economic reality. The literature is already available \cite{cajueiro_possible_2005, cajueiro_role_2008, tabak_topological_2009};
\item Currently, the market for goods is restricted to firms and consumers in the domestic market. However, it could also include firms and governments as buyers (and sellers), enabling analysis of intermediate sectors, as well as foreign buyers allowing the inclusion of an economic measure of exports and trade balance;
\item Although distance is already included in the model, the formula could be sophisticated to effectively include the transport system available in the municipalities that are object of study. As a result, accessibility analysis would be systemically integrated with the rest of the economy, as demand and supply of the transport system (for employment purposes).
\item The process of imposing a limit by time or distance to daily commute would endogenously enable the creation of a system with several regions, making it simple to study urban hierarchy analysis. In such case, the 'employment areas' would be endogenous to the model.
\item Firms and their production technologies, decision-making processes and hiring and firing could be drawn from tacit information specific to a particular firm or sector.
\item The taxation system of this model is simplistic, with only one tax applied to consumption, typically a value-added tax (VAT) levied on the location of the firm. However, note the reader, that the implementation of the Territorial Taxes on property or on income, or changing VAT to be collected at the destination, i.e., at the consumer's place of residence, could be easily implemented. Thus, specific research questions of fiscal interest could be investigated.
\end{enumerate}

Indeed, it is worth mentioning the advantage of modularity within the scope of this work. Using the basic model is possible to detail, build, and expand the model module by module, according to the research needs, while ensuring the evolution of the integration of other processes already implemented and validated.

Anyway, this list is not exhaustive and only fulfills the job of informing the model expansion opportunities, through enhanced feature of this theoretical and methodological proposal.

\section{Final considerations}\label{sec5}

This paper specifies, explains and justifies the steps and processes of the construction of the computational algorithm that prospectively simulates a spatial economy. It adds to the literature on the explicit spatiality of the model, and in achieving a simple model with three markets and conurbated subnational governments. Thus, establishing an actual framework for economic simulation, constituting itself as a public policy tool.

The model has a dynamic real estate market with prices given by the features of the dwelling and its location; a labor market, with matching mechanism between skilled workers and companies; and a goods market with endogenous price adjustment based on stock. The configuration in different subnational governments, one, four or seven differentiated regions allows for explicit spatial analysis.

The results and trends obtained after 1,000 simulation runs indicate that mobility of families among regions is central to the model with impoverished families migrating to poorly serviced places and, therefore, lower real estate prices; and families that are financially well migrating to better quality areas. Therefore, the model with only one region has a less dynamic economy, although more homogeneous, whereas the model with seven regions shows greater dynamism, but also greater heterogeneity and inequality.

The research question that asked whether the change of administrative boundaries and the consequent change of local tax revenue dynamics, in principle, changes the quality of life of the citizens ' can be answered affirmatively. Indeed, administrative boundaries – understood as enclosed area of tax collection over economic base and its investment as collective public services – can alter the quality of life of citizens.
The underlying question faced by this paper is the efficiency of the return of taxes to taxpayers. Is there a spatial, political and administrative configuration that is more efficient? This debate should be further discussed by following research.

Finally, this paper contributes to the methodological framework of economic tools, particularly those flexible and forward-looking, with applied realm to public policies of subnational entities.

\section*{Acknowledgments}

The authors would like to thank the National Council of Research (CNPq/BR).

\bibliographystyle{IEEEtran}
\bibliography{bib_spatial_econ2}

\begin{IEEEbiography}[{\includegraphics[width=1in,height=1.25in,clip,keepaspectratio]{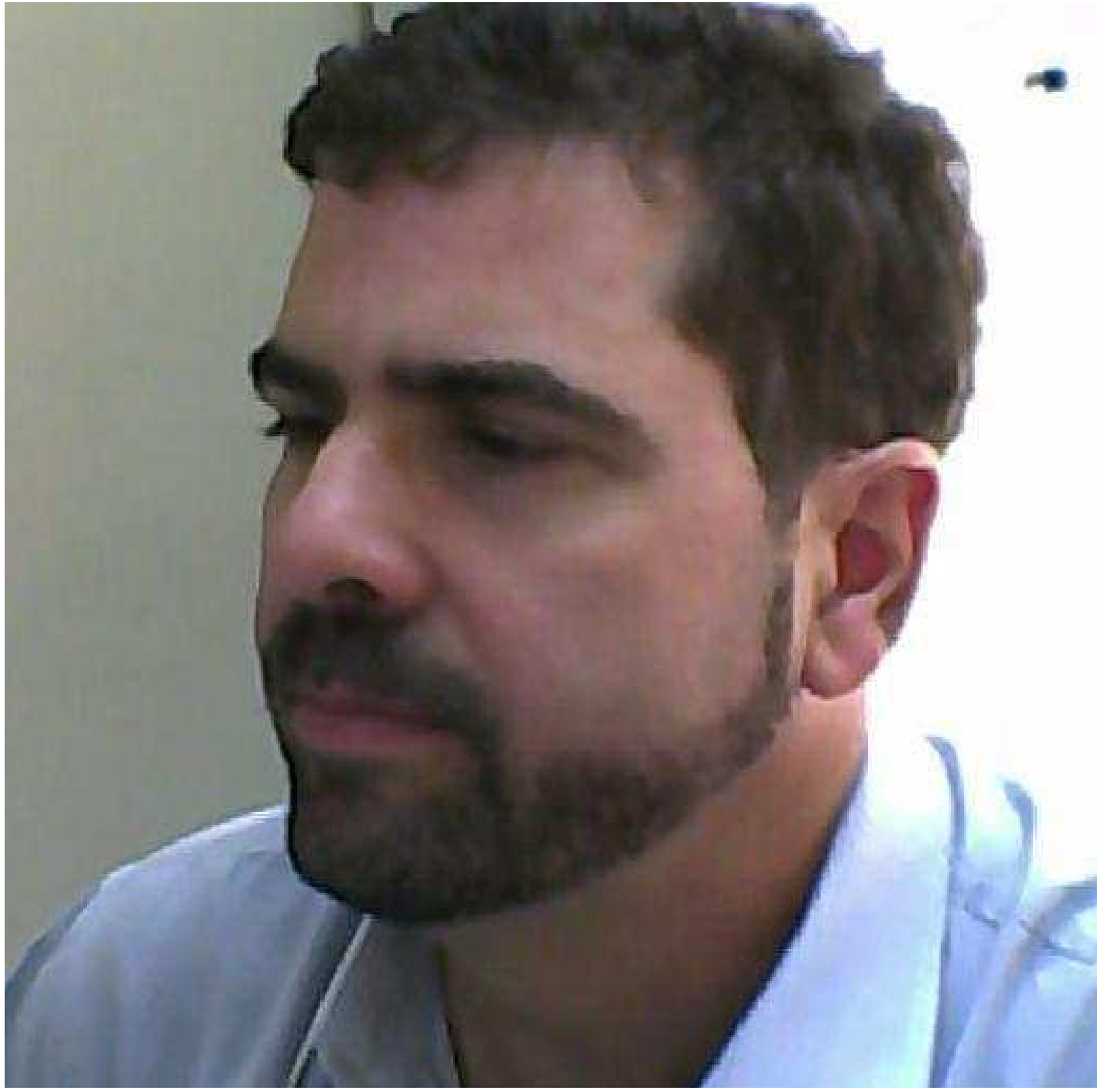}}]{Bernardo Alves Furtado}
	Tenured researcher at government think-tank Institute for Applied Economic Research (Ipea-Brazil), PhD from University of Utrecht and UFMG. Background in urban analysis, economics, and geography. Presently, holds a position as a coordinator at DISET/IPEA and a grant scholarship from CNPq. Former deputy director, editor, professor. Interested in public policy applications of agent-based modeling.
\end{IEEEbiography}
\begin{IEEEbiography}[{\includegraphics[width=1in,height=1.25in,clip,keepaspectratio]{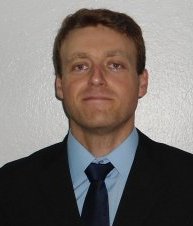}}]{Isaque Daniel Rocha Eberhardt}
	Agronomist engineer from Agronomist School Eliseu Maciel of Federal University of Pelotas - UFPel; Master Degree in Remote Sensing by National Institute for Space Research - INPE; PhD Candidate in Transportation by the University of Bras\'ilia - UNB; and researcher assistant at Institute for Applied Economic Research - IPEA. Main subjects are Remote Sensing, Geographical Information System, Spatial Analysis, R language, Data Mining, Data Analysis, Spatio-Temporal Analysis.  
\end{IEEEbiography}

\appendices\label{Appendix}
\onecolumn

\section{PseudoCode: Real Estate Market}\label{Appendix A}

\textit{Symbol} "\#" \textit{indicates comments about the procedures or functions.} \\

\textbf{Function \textit{Real Estate market}} (families, dwellings, government regions) 

\hspace{1cm}{\# \textit{Select a sample of families to enter the market, given the parameter}}

\hspace{1cm}{Generate list of families on the market,  chosen randomly}

\hspace{1cm}{Create a market dictionary}

\hspace{1cm}{\# \textit{Select vacant houses; update prices and quality for all houses, given QLI in each region}}
		
\hspace{1cm}\textbf{For each dwelling:}
			
\hspace{2cm}{Update prices, given the region}
		
\hspace{2cm}{Update Quality of Life Index}
			
\hspace{2cm}{\textbf{If} dwelling is vacant}
			
\hspace{3cm}{Add it to market dictionary}
			
\hspace{3cm}{Calculate median of families' financial resources}
			
\hspace{1cm}\textbf{For each family that are in the market}
			
\hspace{2cm}{Update family's dwelling value}
			
\hspace{2cm}{\textbf{If} the family has a positive number of members}
		
\hspace{3cm}{Set \textbf{move} equal to False}
			
\hspace{3cm}{\textbf{If} the family's financial resources are below median value}

\hspace{4cm}{\# \textit{Compare all available dwellings, choose the cheapest}}

\hspace{4cm}{Choose a random dwelling (1)}

\hspace{4cm}{For each dwelling:} 

\hspace{5cm}{\textbf{If} dwelling's (2) price is lower than dwelling's (1)}

\hspace{6cm}{Choose dwelling (2)}

\hspace{5cm}{\textbf{Else}}

\hspace{6cm}{Choose dwelling (1)}

\hspace{5cm}{\# \textit{Check the difference between values}}

\hspace{5cm}{\textbf{If} current dwelling is more expensive than the new dwelling:}

\hspace{6cm}{\# \textit{Deduce the difference}}

\hspace{6cm}{Calculate the difference}

\hspace{6cm}{Update families' values}

\hspace{6cm}{set \textbf{move} to True}

\hspace{3cm}{\textbf{Else}}
 
\hspace{4cm}{\# \textit{Compare the available dwellings available, chose the best}}
 
\hspace{4cm}{Chose random dwelling (1)}

\hspace{4cm}{For each dwelling} 

\hspace{5cm}{\textbf{If} dwelling's quality (2) higher than (1)}

\hspace{6cm}{Choose dwelling (2)}

\hspace{5cm}{\textbf{Else}}

\hspace{6cm}{Choose dwelling (1)}

\hspace{4cm}{\# \textit{Check fund availability}}

\hspace{4cm}{If current dwelling price is higher than intended dwelling}

\hspace{5cm}{\# \textit{Deduce the difference}}

\hspace{5cm}{Calculate the difference}

\hspace{5cm}{Update families' dwelling values}

\hspace{5cm}{Set \textbf{move} True}

\hspace{3cm}{\# \textit{Making the move}}

\hspace{3cm}{\textbf{If} \textbf{move} is True}

\hspace{4cm}{Families vacant old dwelling}
 
\hspace{4cm}{\# \textit{New residence}}

\hspace{4cm}{Family register new dwelling, address, value}

\hspace{4cm}{Dwelling register new family}

\section{PseudoCode: Goods Market}\label{sec6B goods market}

\hspace{1cm}{\textbf{Function} Equalize families' funds (families)}

\hspace{2cm}{Distribute equally total resources among family members} \\

\hspace{1cm}{\textbf{Function} Consumption (firms, regions), processed within agent's class}

\hspace{2cm}{\textbf{If} monetary available quantity is positive}

\hspace{3cm}{\textbf{If} resources are lower than unit}

\hspace{4cm}{Consume resources in a random value from 0 to total}

\hspace{3cm}{\textbf{Else}}

\hspace{4cm}{Consumption equals random value between zero and total,} 

\hspace{4cm}{discounted by a $\beta$ parameter}

\hspace{3cm}{\# \textit{Given the size of market, decide among firms,}} 

\hspace{3cm}{\textit{choosing lowest price or closest firm}}

\hspace{3cm}{Create an empty \textbf{market} list}

\hspace{3cm}{Add firms randomly, given market size}

\hspace{3cm}{\# \textit{Choose the firm with lowest price}}

\hspace{3cm}{Create an empty prices' list}

\hspace{3cm}{\textbf{For} each firm on selected \textbf{market}}

\hspace{4cm}{Add prices to the list}

\hspace{3cm}{Select firm listed with lowest price}

\hspace{3cm}{\# \textit{Choosing closest firm}}

\hspace{3cm}{Create an empty distance list}

\hspace{3cm}{\textbf{For} each firm on \textbf{market}}

\hspace{4cm}{Calculate the distance, given consumer's location}

\hspace{4cm}{Add distance of each firm to the list}

\hspace{3cm}{Select from the list, closest firm}

\hspace{3cm}{\# \textit{Choose randomly between closest firm and lowest prices}}

\hspace{3cm}{Choose firm}

\hspace{3cm}{\# \textit{Purchase from chosen firm}}

\hspace{3cm}{Chosen firm processes sales' \textbf{Function},} 

\hspace{3cm}{inputs (amount available to consumption, firm; region); return change}

\hspace{3cm}{Consumer updates cash values, considering spent money and change}

\hspace{3cm}{\# \textit{Utility}}

\hspace{3cm}{Update consumer's utility from consumed value} \\

\hspace{1cm}{\textbf{Function} Sales (consumer's resources, regions), processed within firm's environment}

\hspace{2cm}{\textbf{If} resources are positive}

\hspace{3cm}{\textbf{For} each product in inventory}

\hspace{4cm}{\textbf{If} quantity is positive}

\hspace{5cm}{Purchased quantity equals available resources divided by price}

\hspace{5cm}{\# \textit{Verifying if available quantity is enough}}

\hspace{5cm}{\textbf{If} demanded quantity is higher than offered quantity}

\hspace{7cm}{Calculate amount spent, given quantity and prices}

\hspace{7cm}{Deduce quantity from firms' inventory}

\hspace{7cm}{\# \textit{Taxes and balance}}

\hspace{7cm}{Deduce spent amount from agent}

\hspace{7cm}{Pay firms, deducing owed taxes, given rates}

\hspace{7cm}{Add collected taxes to each region}

\hspace{7cm}{\# \textit{Quantity sold}}

\hspace{7cm}{Generates statistical information}

\hspace{1cm}{Return change, if necessary}

\section{PseudoCode: Labor Market}\label{sec6C labor market}

\textbf{class Matching} \\

\hspace{1cm}{\textbf{Function} Firms offer position}

\hspace{2cm}{Create dictionary for firms offering positions}

\hspace{1cm}{\textbf{Function} Candidates application}

\hspace{2cm}{Create dictionary with agents that fill prerequisites}

\hspace{1cm}{\textbf{Function} Matching}
	
\hspace{1cm}{\# \textit{Rank candidates by qualification}}

\hspace{2cm}{\# \textit{Matching}}

\hspace{2cm}{\textbf{While} both candidates' list and firms' list contain elements}

\hspace{3cm}{Choose random firm (1) among those on dictionary}
 
\hspace{3cm}{\# \textit{Higher qualification of candidate}}

\hspace{3cm}{Choose random candidate (3)}

\hspace{3cm}{\textbf{For} each candidate}

\hspace{4cm}{\textbf{If} candidate qualification (4) is higher than candidate (3)}
 
\hspace{5cm}{Select candidate (4)}

\hspace{4cm}{\textbf{Else}}

\hspace{5cm}{Select candidate (3)}

\hspace{3cm}{\# \textit{Firm's closest candidate}}

\hspace{4cm}{Choose random candidate (5)}

\hspace{4cm}{\textbf{For} each candidate (6)}

\hspace{5cm}{\textbf{For} each firm (7)}

\hspace{6cm}{\textbf{If} firm's distance is lowest}
 
\hspace{7cm}{Select candidate (5)}

\hspace{6cm}{\textbf{Else}}
 
\hspace{7cm}{Select candidate (6)}

\hspace{3cm}{\# \textit{Choose randomly between best candidate or the one who lives closest}}

\hspace{3cm}{Choose candidate}

\hspace{3cm}{Firm hires, candidate register job}

\hspace{3cm}{\# \textit{Delete candidate and firm from the dictionary}}

\hspace{3cm}{Delete firm that hired from dictionary}

\hspace{3cm}{Delete chosen candidate from dictionary}

\twocolumn

\section{ODD Protocol}\label{odd}

The model description follows the ODD (Overview, Design concepts, Details) protocol for describing individual – and agent-based models \cite{grimm_standard_2006, grimm_odd_2010}.

\begin{enumerate}
\item{Purpose}

The model is a first exercise to observe the economy in its full spatially explicit environment, its markets, and main agents in order to capture taxes mechanisms and their effects as a means to enable public policy evaluation ex-ante. The model falls within the simple category and it is a loosely expansion of Lengnick \cite{lengnick_agent-based_2013} and Gaffeo \cite{gaffeo_adaptive_2008} with the introduction of spatially bounded government regions. Proven its validity, Brazilian intricate tax system can be inserted into the model and subnational development in the medium term could be analyzed. The main hypothesis is that for a given labor market area where citizens commute daily; making changes to political administrative boundary impacts citizens quality of life. The results of the model suggest that it may be the case.

\item{Entities, state variables, and scales}

The model contains classes of agents, families, dwellings, firms, and government, along with accessory classes. 
IDs. All members of all classes have their own unique identification (IDs). Agents keep their current workplace ID, family and dwelling IDs. Families keep track of all their members and current dwelling. Dwellings knows which family is currently hosting, if any, their government region ID. Firms knows all their employers IDs and its government region ID. Government is passive and money is transferred directly from consumers at the buying moment. Each government region has its own ID.

\textbf{The agents} have age (in years), qualification (in years of study), utility and money attributes along with family and dwelling identification. They also have processes to update money balance and consumption procedures. Agents buy and work in firms, are members of families and can move among dwellings along with their families.  

\textbf{Families} are groups of agents. Consumption money is always equally divided among family members. Families move among dwellings. They register the current dwelling address and value. 

\textbf{Dwellings} are fixed in space, have prices (proportional to size and Quality Index of the government region), size and quality and addresses. Quality and prices are updated monthly, given that the Quality Index at the government region has been updated.  

\textbf{Governments} are within sets of pre-defined boundaries. They have a Quality Index and collect taxes deducing amount from firms when selling to costumers within their territory. They invest all treasure into Quality Index updates. 

\textbf{Firms} have one product (with price and quantity), monthly balance, profits, addresses (x, y), and constantly knows its employees. They are fixed in space and process sales, production (with product quantity update), hire and fire decisions, and make employers payment. 

\textbf{Temporal extent}. The model was designed to run for 20 years, but there is no strict limit applied. It runs in terms of days, months (of 21 days), quarters and years, following Lengnick \cite{lengnick_agent-based_2013}. In this configuration 5,040 days represent 20 years.

\textbf{Spatial extent}. The spatial boundaries are determined exogenously through parameters. We used -10, 10 on horizontal and vertical axis with location within the square in float precision.

Along with the main classes, accessory classes include a system of communication that holds the labor market procedures; statistics, output, plotting, main, control (that iterates over simulations), generator that creates the instances of agents, families, dwellings and firms before simulation; parameters; space and time iteration. A products class was also built to facilitate new products development. However, in the current model only one product is in effect.

\item{Process overview and scheduling}

The model runs in a discrete mode fully using the Object-Oriented Programming (OOP) paradigm. A time schedule procedure is described in \autoref{sec2}. The pseudocode of the main processes are also available as Annexes \autoref{Appendix A}, \autoref{sec6B goods market}, and \autoref{sec6C labor market}. 

The model was designed for a Python 3.X environment. Thus, it makes full use of classes, their variables and methods with variable updating being processed when methods are called. In order to run one simulation step, instances of the classes are created to make sure that objects are updated correctly. According the description of item 2, we have the following instances: my\_agents, my\_regions, my\_houses, my\_firms, my\_journal (communications), my\_parameters, my\_simulation (TimeControl). R is used to process the results and produce plots.

\item{Design concepts}

\textbf{Basic principles}. The model is an extension of some principles from Lengnick \cite{lengnick_agent-based_2013} and Gaffeo et al. \cite{gaffeo_adaptive_2008} with many adaptations. Given the lack of simple models \cite{van_der_hoog_production_2008}, some processes are new implementations. Mainly these processes are the markets of goods, labor and houses and the decision processes of firing and hiring, setting of goods prices and wages, following earlier literature \cite{bergmann_microsimulation_1974, blinder_inventories_1982, blinder_sticky_1994, michaely_corporate_2012}.

\textbf{Prices}. Firms rationalize on prices given their current stock. When quantity produced is below a given parameter threshold, firms raise prices by some parameter percentage, i.e. price is given by cost but when there is enough demand a mark-up value is added to price. Otherwise, when quantity is above that same parameter threshold price is given by cost. 

\textbf{Wages}. Firms pay a constant wage plus a qualification additional. 

\textbf{Goods market}. Firms offer their products with given prices. Consumers choose from a subsample of firms; which is determined by a parameter of the model. When considering to purchase, consumers randomly decides from firms that either have the minimum price \cite{mankiw_principles_2011} or is the closest from its residence \cite{fujita_spatial_1999, losch_economics_1954}. 

\textbf{Labor market}. Candidates of age with a given qualification offer themselves repeatedly on the market. Firms offer posts. Matching happens between most qualified employee and either (randomly) or closest firm. 

\textbf{Hiring and firing}. Firms decide on hiring or firing periodically, given an exogenous parameter. Typically, once every four months. When profit or cashflow is positive or the firm has no employees, they offer one post. When profit is negative, they fire an employee. Hiring follows the process described in the labor market. The employee to be fired is chosen randomly from the pool of employees of the firm.

\textbf{Housing market}. Families enter the housing market periodically, following an exogenous parameter \cite{ibbotson_world_1985}. Once in the market, their decision to move is for more quality, when families have wealth above the median of all families; or to move for cheaper houses and capitalizing on the difference, following a general model of urban economics \cite{brueckner_structure_1987, dipasquale_urban_1996}. In the first case, for each family in search, the available houses are ranked in order of quality. If the family’s wealth plus the value of the current house is enough to best house, the transaction and the move occurs. In the second case, each family goes for the cheapest house.

\textbf{Emergence}. As output of the model, typical indicators of the economy are produced. We believe many of the results to be endogenous and robust on the parameters of the model. Built-in the model is the idea of economies of agglomeration and disagglomeration \cite{fujita_spatial_1999} and urban economics in general\cite{brueckner_structure_1987, dipasquale_urban_1996}. That is, the center (or the region) where there is a concentration of firms that are performing well tend to attract agents. Simultaneously, rent prices rise and such process expels families to poor suburbs. However, the central region is an endogenous result of the model. 

\textbf{Adaptation and Learning}. Agents, families or firms do not adapt in the sense that they change the process of decision-making. However, the looseness of families in respect to dwelling attachment along with the dynamics of price changing implies that families have to financially adapt to constant changing environment. 

\textbf{Objectives}. Families' objective is either try to move to better quality places or have enough resources to keep consuming. Their ‘success rate’ is measured by their members’ utility, which is an indicator that cumulatively measures their actual consumption. Firms’ objective is to increase in size and keep on hiring and producing more and more. However, when out of equilibrium, objective changes to restore financial health (firing employees). Government (implicit) objective is to increase Quality Index. 

\textbf{Prediction}. Decisions of all agents are based on cross-sectional information and do not try to infer the future. The model as a whole is intended as a comprehension mechanism and as a provider to the specific questions posed. 

\textbf{Sensing}. Sensing is global for agents when looking for jobs. When consuming, sensing is restricted to the given number of firms defined (parameter) as its private market. Dwelling price and quality changing mechanism is proportional to the observed at their region; which in turn is proportional to the sales of firms acting in the region. 

\textbf{Interaction}. Interaction happens competitively at the three designed models. However, agents and firms interact directly only in the sense that they may be excluded from the selection process (of hiring or consuming), given that they have lower qualification, are offering lower salaries or expensive products. Implicitly, agents at the same region share the same Quality Index. Thus, families benefit from high quality dwellings that is given by profitable firms in the region.

\textbf{Stochasticity}. Stochasticity plays an important role at this version of the model. All population of agents, families, firms, dwellings is generated from a random process. Random decisions between two alternatives occur when deciding either for closest or cheapest product and for most qualified or closest living employee. Further, there is a random process when the agent decides the amount for consumption monthly and when the firm makes its firing decision.

\textbf{Collectives}. The only collective in the present model are the families. They have been described as a class above. 

\textbf{Observation}. A number of statistics are collected on a monthly basis. However, they do not interfere endogenously, except for the average families’ wealth (interferes on moving decision). The indicators available are absolute production sold (GDP), unemployment, average number of employees per firm, average utility of agents, average prices, average firms' balance, sum of firms' profit, Gini inequality index (based on families' average utility). Firms individually calculate and use their profits endogenously. Every three months they record their total balance and then they calculate next months’ profit in relation to the recorded value.

\textbf{Parameters}. The following parameters are requested from the modeler at every simulation (\autoref{sec2}). 

\textbf{Iteration}. Given the artificialness of the population and the stochasticity described, the model was run 1,000 times for each regional configuration (one, four or seven regions) and the results are presented in terms of distributional statistics. 

\item{Initialization}

At time 0 of the simulation, a number of processes has already run and will not run again (\autoref{sec2}). Given that this model is an artificial test aimed at scrutinizing (and proposing) the model itself and allowing only a hinted indication of public policy, the population is always a different one for each different run, given the parameters discussed. It is our plan to apply the model to a metropolitan area with fixed given population.

\item{Input data}

The model does not use input data, as it is.

\item{Submodels}

There are no submodels in this version.

\end{enumerate} 

\end{document}